# The Naive Extrapolation Hypothesis and the Rosy-Gloomy Forecasts

Vasileios Barmpoutis

Harvard University, Kennedy School

## Abstract[*]

I study the behavior and the performance of the long-term forecasts issued by financial analysts with respect to the Extrapolation Hypothesis. That hypothesis states that investors, extrapolating from the firms' recent performances, are too optimistic about growth and large firms and too pessimistic about value and small firms. I find that the forecasting errors are higher for the growth firms and large firms, thus providing support for the Extrapolation Hypothesis. However, in addition to the rosy picture of the growth and large firms, the forecasts of the value and small firms are not so gloomy in many cases. My analysis also reveals that expectations move together for all categories of book-to-market and all sizes of firms. I proceed by investigating some common factors that may influence analysts' long-term forecasts, including co-movement and excessive optimism. I find that macro factors beyond a firm's recent performance may influence the formation of expectations.

## I. Introduction

It is a well-established fact that value strategies, such as buying cheap stocks using book-to-market (B/M) and price-to-earnings ratios as measures, earn superior returns over sufficiently long periods. However, the reasons for this are controversial. Two competing theories have been offered to explain the phenomenon. Fama and French (1992) explain the higher returns of value stocks by the higher risks involved. In contrast, Lakonishok, Schleifer, and Vishny (LSV) (1994) point to behavioral explanations. More specifically, LSV (1994) propose the hypothesis that investors naively project the current performance of firms into the future. But, in reality, most firms cannot sustain their previous growth rates, and eventually their performance means revert. Growth firms slow down, and value firms improve their performance. The stock market realizes the situation with some lag, and the returns adjust accordingly. La Porta (1996) finds evidence that the practice of buying stocks with low forecasted growth and selling stocks with high forecasted growth earns superior returns. Also, La Porta, Lakonishok, Shleifer, and Vishny (1997) find that the reactions to earnings announcements are more positive for value stocks than for growth (glamour) stocks for a period of 5 years after the portfolio formations. Doukas, Kim, and Pantzalis (2002) sort firms into book-to-market and size quintiles and use the EPS forecasts of financial analysts to examine the forecasting errors with respect to the classification of firms. They find that the beginning-of-the-year earnings forecasts are actually more optimistic for value and small stocks, contrary to the predictions of the Extrapolation Hypothesis.

[*] I am grateful to François Degeorge, Roni Michaely, Alexander Wagner and Richard Zeckhauser for assistance, comments and suggestions. The financial support by the Swiss Finance Institute, the FNS - ProDoc – PDFMP1_126390 project and a Swiss National Funds fellowship made this work possible and are gratefully acknowledged. All errors are mine. Please send all correspondence to vasileios_barmpoutis@hks.harvard.edu.

In this study, I present evidence that the long-term forecasts of financial analysts are indeed more optimistic for the growth and large firms, providing support for the Extrapolation Hypothesis. In that sense, this study's results are close to those of La Porta (1996), who also uses analysts' long-term growth forecasts. However, La Porta doesn't test the Extrapolation Hypothesis directly by sorting the firms into book-to-market and size portfolios and then testing their long-term performance, which is the procedure I follow in this study. More specifically, I find that errors in forecasted growth are greater for the low book-to-market firms and for large firms. Also, for most cases during the sample period, the forecasting errors are positive for all categories of book-to-market and for all sizes of firms. In other words, analysts overestimate the growth prospects of all firms, but they are particularly overly optimistic about the outlook of growth and large firms. The only category of firms that exhibits negative forecasting errors is the very small value firms.

Another property of long-term forecasts is that analysts' expectations move together, that is, they increase and decrease together for all categories of book-to-market and for all sizes of firms. I postulate that there must be some macro factors that force analysts to update their forecasts for all firm categories at the same time. I explore the phenomenon by investigating the relationships between analysts' expectations and a series of possibly influential factors. I find that analysts' forecasts are mainly influenced by mean firm age and by mean firm performance during the previous year. However, analysts are not particularly influenced (or they are influenced in the wrong direction) by overall market returns, the firm's specific returns, GDP growth, GDP growth forecasts issued by the Survey of Professional Forecasters (SPF), and the firm's quarterly earnings. Moreover, analysts' forecasts are not in agreement with, and are overly optimistic as compared to, the corporate profit forecasts produced by SPF. Further, I provide some preliminary evidence that long-term forecasts are influenced by the number of IPOs and the total factor productivity (TFP). The influence of IPOs is not limited to new firms just entering the stock market, but also includes the firms already listed. Overall, it seems that optimism among analysts is not the result of only the performance of individual firms, but also the result of economy-wide developments.

In general, the results point to the fact that the inferior performance of growth stocks seems to be the result of the previous excessive optimism, as described in LSV (1994) and others. Also, the study provides some evidence that the extrapolation of fundamentals (for example Barberis, Schleifer and Vishny (1998), Fuster, Herbert and Laibson (2011)) may play some role in explaining the observed patterns in the stock market.

The paper is organized as follows. Section II describes the data. Section III examines the Extrapolation Hypothesis. Section IV examines the common movement of expectations and errors. Section V investigates some factors that contribute to the common movement. Section VI concludes.

## II. Data

The data are drawn from IBES, CRSP, and Compustat. I use the IBES summary long-term forecasts file from 1981 to 2011, the CRSP monthly and daily files from 1981 to 2011, the CRSP events file from 1955 to 2011, and the Compustat annual fundamentals file from 1981 to 2011. For the analysis related to long-term forecasts, I use the intersection of firms among the CRSP monthly file, the IBES summary long-term forecasts file, and Compustat. For the analysis of the firms' accounting data, I use the intersection of the CRSP events file and the Compustat annual and quarterly fundamentals file.

I create the intersection of the IBES long-term summary file and the CRSP monthly file, using the connection table provided by WRDS. I also use the procedure proposed by WRDS to connect CRSP and Compustat. I use the CRSP identifier (permno) as a bridge to connect IBES and Compustat files.

Descriptive statistics for the IBES summary long-term forecasts file (intersection with CRSP) are presented at Figures [1] and [2]. Figure [1] reports the number of firms per year for the 1981-2011 period. Figure [2] presents the number of estimates per year for that same period.

The CRSP monthly file includes 21,120 firms from three exchanges (NYSE, AmEx, and NASDAQ). The CRSP stocks are assigned to five size quintiles. The quintile breakpoints correspond to NYSE quintiles, and each stock is assigned to each portfolio at the beginning of each year, based on its market equity at the end of the previous year. Each stock stays in the portfolio for a year.

For calculating the book-to-market of each firm, I use accounting data from the Compustat annual fundamentals file. I calculate the book value as the sum of assets, deferred taxes, investment tax credit, and convertible debt, minus the preferred stock and total liabilities. I calculate the market value at the end of the current calendar year, and then I calculate the book-to-market ratio. I use NYSE breakpoints to calculate the quintile of the book-to-market values each year. The quintile to which a firm belongs is the same for the entire next year; the quintiles calculated in 12/2001 are used for the entire year 2002.

Figure [3] presents the number of firms per book-to-market quintile in the intersection between the IBES summary long-term forecasts file and the CRSP monthly file. Figure [4] presents the same information with respect to the size quintiles.

# III. Extrapolation Hypothesis

The Extrapolation Hypothesis states that investors' expectations are too high for growth firms (low book-to-market firms) relative to value firms (high book-to-market firms). Also, according to Fama and French (1992), the high returns of value stocks are related to small rather than large firms, making this another basis for extrapolation. Moreover, investors extrapolate to the future based on firms' recent performances.

Doukas et al. (2002) use analysts' earnings forecasts to test directly the Extrapolation Hypothesis and find no support for it. Specifically, they sort the stocks into book-to-market quintiles; they find that the stocks which belong to the highest book-to-market quintile, the value stocks, exhibit the largest forecast error, while the stocks of the first quintile, the growth stocks, exhibit the smallest forecast error. According to Doukas et al. (2002), this finding provides evidence that investors are more optimistic about value stocks (assuming that analysts' forecasts proxy for the investors' expectations), and less about growth stocks, contrary to the situation described by the Extrapolation Hypothesis. Doukas et al. repeat the sorting procedure with respect to the size of firms. They sort the stocks into five size quintiles; they find that the smaller firms have higher forecast errors than the larger firms. They interpret this as a finding that investors do not overestimate the earnings of larger firms, contrary to what the Extrapolation Hypothesis implies.

Instead of using quarterly and yearly earnings per share (EPS) forecasts, I use analysts' long-term growth forecasts to test the Extrapolation Hypothesis. An advantage of long-term growth forecasts is that they provide expected growth rates, so I can compare not only analysts' errors but also the expected growth rates per se among firms. Note that using earnings forecasts to compare errors across firms has two difficulties. The first is that analysts seem to cooperate with management; they forecast EPS in such way that management can beat the forecasts (Degeorge, Patel, and Zeckhauser, 1999). The other is that dividing the errors in order to facilitate comparison across different firms, for example, by price, may alter the distribution of errors across firms (Cheong and Thomas, 2011).

Long-term growth forecasts correspond to an average annual increase in operating earnings over a company's next full business cycle. According to IBES, these forecasts refer to a period of three to five years. I use a value of four years as an intermediate value. In order to define long-term forecast errors, I have to attribute the forecasts to an accounting quantity. I use the operating income before depreciation. The IBES manual gives the following description for long-term forecasts:

*"Long-term growth rate forecasts are received directly from contributing analysts; they are not calculated by Thomson Reuters. While different analysts apply different methodologies, the Long-term Growth Forecast generally represents an expected annual increase in operating earnings over the company's next full business cycle. In general, these forecasts refer to a period of between three to five years."*

According to Compustat (p. 365 of the Compustat manual), the operating income before depreciation is calculated as Sales (Net) minus Cost of Goods Sold and Selling, General and Administrative expenses before deducting Depreciation, Depletion, and Amortization. The definition takes into consideration both sales and cost of goods sold, so it is a measure close to the "operating earnings" suggested by IBES.

I define two types of errors, simple and relative. The simple forecast error is defined as:

$$FE_{i,t} = Forecasted\_OIBDP_{i,t-4} - OIBDP_{i,t}$$

and the relative as

$$RFE_{i,t} = \frac{(Forecasted\_OIBDP_{i,t-4} - OIBDP_{i,t})}{OIBDP_{i,t}}$$

with $OIBDP_{i,t}$ the operating income before depreciation for firm i at quarter t and with $Forecasted\_OIBDP_{i,t-4}$ the operating income forecasted four years previously for the firm i.

In other words, I check the operating income, before depreciation, of the current quarter and of the same quarter four years previously. By doing so, I create a time series of forecast errors for every firm.

The relative forecast error is superior since it is independent of the magnitude of the operating income. Since the numbers are not per share, the problem of a very small denominator is largely mitigated.

The choice of operating income as the chosen accounting variable comes with two drawbacks. First, there are some missing values: 157,119 of the 683,300 potential observations are missing from the Compustat Fundamentals file. Second, the measure is too volatile. The problem introduced by high volatility is that, for example, a spike in a given quarter will lead to the erroneous conclusion of better-than-expected performance, and vice versa for a drop. I try to mitigate the first problem by replacing missing values with the average of the quarters before and after the missing one. In so doing, I increase the sample size by 7,344 observations. To address the second problem, I use a moving average of the current and the previous four quarters of a firm's operating income. Figure [5] presents the case of Microsoft's operating income as an example. The employed procedure seems to reduce effectively the inter-temporal volatility and also follows the trend of Microsoft's operating income.

**III.A. Book-to-Market**

In order to test the Extrapolation Hypothesis for growth and value firms, I sort the stocks in book-to-market quintiles, and I examine the median forecast error. I use median errors in order to avoid the influence of outliers. Use of medians for long-term growth forecasts is also advised by I/B/E/S. Table [1] presents the results for the simple error, the relative error, and the expectations for each quintile for the whole sample[1]. The errors, both absolute and relative, are positive for all book-to-market quintiles, pointing to analysts' optimism. However, there is significant dispersion among the different quintiles. At the first quintile, corresponding to the growth firms, the relative error is 21%. This means that the operating income before depreciation after four years is overestimated by 21%. The overestimation for the high book-to-market firms is only 1.26%, which is negligible compared to the error of the growth stocks. The errors for the three middle quintiles are 14.5%, 6.9%, and 3.7%; that is, the error decreases monotonically for higher book-to-market stocks. Notice also that the growth expectations for growth stocks are higher than the growth expectations for value firms; 20% is the median growth expectation for growth firms, versus 10.68% for the value firms. The growth expectations also decrease monotonically with the book-to-market quintile.

---
[1] Because the comparison is between forecasts and the realization four years later, the sample of forecasts for the case of computing the forecast bias is until 2007, but the actual results reach 2011. So, sample is until 2011 because of the use of the 2011 actual results and not of forecasts issued in 2011. The full sample of forecasts until 2011 is used in other sections of the paper.

I follow the procedure used by Doukas et al. (2002), and I examine the firms by exchange. The results are presented in Table [2]. For the NYSE firms, the results are the same as above, namely high errors and expectations for the growth firms and low errors and expectations for the value firms. Both errors and expectations decline monotonically. For the AmEx, the errors for the value firms are higher than the errors for the growth firms, but only by 3%. For NASDAQ, the errors are higher for the growth firms than for the value firms (28% difference); the decline in errors is monotonic; and the errors of the NASDAQ value stocks are negative, meaning that analysts underestimate the firms' prospects.

A valid criticism could be that the forecasts project too far into the future, so errors are inevitable. For this reason, even if the argument doesn't address the issue of consistent optimism, I examine the performance of the firms two years after the release of the forecast, using the fact that the forecasts refer to average annual growth. The results are presented in Table [3]. The magnitude of errors is smaller; however the ordering is the same. The errors for the growth firms are higher than the errors for the value firms: 6% versus 0.4%. The errors still decline monotonically. The picture is the same even if I divide the stocks by exchange. Those results are presented in Table [4]. For all three exchanges, the growth firms' forecasts exhibit higher errors than those of the value firms. For the NYSE and NASDAQ, the errors decline monotonically by the book-to-market quintile. Also, the errors of the last three quintiles are negative.

The above results seem to confirm the Extrapolation Hypothesis. The forecasts for the low book-to-market firms are more optimistic than the forecasts for the value firms.

**III.B. Size**

I now examine the Extrapolation Hypothesis with respect to firm size. I sort the stocks into size quintiles, and I examine the median forecast error. Table [5] presents the results for the simple error, the relative error, and the expectations for each quintile for the whole sample. As in the case of the book-to-market sorting, the errors, both simple and relative, are positive for all size quintiles, pointing to general optimism. Again, there is significant error dispersion among the different quintiles. For the small firms of the first quintile, the relative forecast error is negligible, while, for the large firms of the fifth quintile, the median relative error is 13%. The growth forecasts are higher for the smaller firms and decrease monotonically by size.

The results remain the same if the firms are examined by exchange. The results are presented in Table [6]. For NYSE, the results mirror the results of the whole sample. For AmEx, the results are qualitatively the same, but the median error of the large firms is 87%. This is not because of higher expectations, since the expectations for all AmEx firms are the same, around 15%. For NASDAQ, the relative error is marginally negative for smaller firms and positive at 18% for the largest firms. The forecasted growth rates stand at 17-18% for all NASDAQ firms.

I examine again the performance of the forecasts two years after their release. Table [7] presents the results. The overall picture remains the same. Smaller firms have small and negative errors, and large firms have larger errors. The conclusions are the same, even if categorize the firms by exchange (Table [8]). The relative forecast error is 2% for the smaller NYSE firms and 5% for the larger firms. In the case of AmEx, it is 1% for the smaller ones and 26% for the larger ones. For NASDAQ, the results are -1% for the smallest and 5% for the biggest.

In general, it seems that the relative errors of the forecasts for the large firms are larger than the relative errors for the smaller firms.

### III.C. Book-to-Market and Size

Finally, I sort stocks by book-to-market and size. The results of the relative errors are presented at Table [9] and confirm the previous results. The small and high book-to-market firms have negative relative errors (-8%), meaning that their prospects are underestimated (providing support to the Extrapolation Hypothesis), while the big growth firms have relative errors of 21%. For all book-to-market quintiles, the largest errors are observed at the fourth size quintile, while for all size quintiles the largest errors are observed at the first book-to-market quintile. As a conclusion, the results confirm one of the predictions of the Extrapolation Hypothesis: analysts are optimistic for growth firms and large firms, but the results are quite disappointing. Growth firms miss the forecasts by around 20% and large firms by 13%. On the other hand, value and small firms miss the forecasts by only 1% and 0.6% respectively.

### III.D. Comparison with the Results of Doukas et al. (2002)

The results of the previous three subsections regarding the Extrapolation Hypothesis are in contrast to the results presented by Doukas et al. (2002). It is true that I focus on operating income, while the study by Doukas at al. (2002) focused on earnings. I focus on a longer horizon, and my sample covers a longer period. However, it could be useful to investigate possible reasons behind this discrepancy.

A reproduction of the results of Doukas et al. (2002) using data available up to 2011 reveals that their results are still valid. Also, they are very robust to denominators other than the stock price.

I turn to the behavioral aspects of the production of earnings forecasts in order to resolve the disagreement between my results and the results of Doukas et al. (2002). Degeorge et al. (1999) provide strong evidence of the importance of thresholds in producing earnings estimates. They identify three thresholds. The first is the zero threshold; that is, positive earnings are preferable to negative earnings. The second is the sustained performance; that is, it is preferable to report earnings equal to or higher than the previous year's. The third is to beat earnings forecasts. If analysts also use thresholds, it can be expected that, for the first forecasts at the beginning of the year, the first and second thresholds will be far more important, since the actual earnings at the end of the year will be unknown even to the management. So, according to this hypothesis, analysts will tend to forecast earnings that are positive and in excess of the previous FYE results. The fact that analysts will report earnings favorable to management is also supported by Lim (2001). According to this, it is useful to examine the earnings forecasts at the beginning of the current financial year with respect to the actual EPS of the previous year. Table [10] reports such results for the forecasts issued eight months or earlier before the FYE. Panel A concerns the results of the B/M quintiles. The results show that, with respect to the previous actual results, analysts increase their forecasts more for the growth firms than for the value firms. Panel B concerns the size quintiles. Those results show that analysts increase their forecasts approximately the same for all quintiles if the median is used, or more for the large firms if the mean is used. Overall, the EPS growth (with respect the previous year's results) forecasted at the beginning of the financial year is more optimistic for the growth firms and approximately the same for all size quintiles.

## IV. <u>Time Series of Relative Errors and Common Movement</u>

The tables in section III provide evidence that the Extrapolation Hypothesis is valid for the 1981-2011 period, based on the financial analysts' long-term growth forecasts. However, the results may apply only to specific years or periods, not to the whole 1981-2011 period, and may extend to the whole period through the process of averaging. In this section, I address the concern that the above results for book-to-market and size classifications may be driven by specific years or periods. I examine first the case of book-to-market classification. In the case under consideration, I have five time series, one for each book-to-market quintile, with one data point for each quarter. The results are presented in Figure [7]. The results confirm the conclusions of Tables [1] through [4]. The relative errors for the growth firms are always higher than the errors for the value firms. The errors across quintiles seem to move together. Also, in most of the sample, the errors decline monotonically by the B/M quintile.

I also examine the possibility that the observed behavior of size-quintile simple and relative errors in Table [5] is the result of certain years. The time series of relative errors for all size quintiles are/ presented in Figure [9]. The results confirm the conclusions of Tables [5] through [8]: the relative errors of the larger firms are always higher than the relative errors of the smaller firms for the whole period. The errors also seem to move together, as in the case of firms sorted by book-to-market. In many periods, the errors of the firms of the fourth quintile are higher than the errors of the firms of the fifth quintile.

Next, I employ the above methodology to investigate whether the common trends of errors have counterparts at the forecasted growth rates. If the same trends are found also in the expectations, it would mean that the errors of the forecasts are not driven only by unexpected subsequent developments that influenced certain firms in certain ways. The results for the book-to-market quintiles are presented in Figure [6]. It is evident that there is strong covariance among the forecasts of various quintiles. They all start increasing at the beginning of the 1990s, peak around 2000, and then drop. The correlations of the forecasts are reported in Table [11] Panel A. They are all above 0.5 and significant at the 1% level. The correlations of relative errors among the book-to-market quintiles are reported in Table [11] Panel B. They are all positive and economically significant.

I repeat the above procedure for the case of the size quintiles, presented in Figure [8]. The correlations among the expectations (Table [12] Panel A) of the size quintiles are all above 0.59, implying that all rise and fall together. The correlations between relative errors and expectations are presented in Panel B; they are economically and statistically significant.

As a conclusion, Figures [7] and [9] suggest that the Extrapolation Hypothesis applies to the whole period of 1981-2011. Moreover, those figures suggest that the relative errors for the different book-to-market and size quintiles move together. Figures [6] and [8] provide evidence that the long-term forecasts of all quintiles also move together.

## V. **Possible Factors Explaining the Common Movement**

The previous two sections established the fact that there is evidence that the Extrapolation Hypothesis is valid with respect to the long-term analysts' forecasts and for the whole 1981-2011 period. Moreover, the relative errors and expectations of the various book-to-market and size quintiles move together. In this section, I examine some factors that might influence the formation of long-term expectations and contribute to their common movement, in order to illuminate the factors that influence analysts and lead them to overestimate the future performance of growth and large firms.

### V.A. Overall Market Returns

The rationale behind examining market returns is that high or low market returns may influence financial analysts' forecasts. This may happen in at least two ways. First, by taking into consideration that markets are forward-looking, high market returns may signal improved prospects for the economy and, therefore, for firms. Second, high market returns may create euphoria among analysts, who then believe that the market and the economy as a whole will continue having sustained growth. This euphoria could translate into higher expectations for firms' growth.

As a first step, the correlations between the SP 500 and the B/M and size quintiles are calculated. Table [13] presents the Pearson correlations between each B/M and size quintile and the SP 500 index. The results remain the same if the Spearman correlations are used. For the whole sample (1981-2011), the correlations are positive and economically and statistically significant for all quintiles. However, the results change if the sample is divided into two periods. The correlations for the current quarters between forecasts and SP 500 are very small for all quintiles (in the range of 5%); they are economically and statistically insignificant for the first subsample from 1996 to 2011. By contrast, they are positive and significant economically and statistically (with the exception of the second quintile) for the 1981-1995 period.

The results for the size quintiles are presented at Table [13] Panel C. The results for the whole sample indicate that the correlations are positive and economically significant for all size quintiles. For the years after 1995, only the fourth and fifth quintiles exhibit any significance; and, economically, the significance is rather reduced. For the years before 1996, the correlations for the first four quintiles are economically and statistically significant.

In order to further investigate and quantify the relationship between the SP 500 and the long-term forecasts, I run the following regression for each B/M and size quintile:

$$D.forecast = a \times LD.forecast + b \times Ret(SP500) + c \times L.Ret(SP500) + d \times SP500 + e \times L.SP500 + \varepsilon, \qquad (1)$$

where *L.* indicates the lagged value, *D.* the difference, and *LD.* the lagged difference.

I run the regression for the whole sample and for each of the two periods after 1995 and before 1996. I include both the returns and the actual level of the SP 500, since they may also convey information to analysts; for example, a new record for the SP 500 may increase optimism. The results for the B/M quintiles are presented at Table [14]. For the whole sample (Panel A), the previous quarter return of the SP 500 is significant for growth firms. The current return and the previous quarter level are significant for the second quintile, while for the other three quintiles there is no significant factor. The $R^2$ is low for the first two quintiles and close to zero for the other three. For the period after 1995 (Panel B), the previous quarter return is again significant for the growth firms and for the firms of the second quintile. For the period before 1996 (Panel C), the only significant coefficients are those of the second quintile.

Table [15] presents the results for the size quintiles. Previous quarter returns are significant for the medium and large firms (third, fourth, and fifth quintiles). Current returns and current and past levels are significant for the smallest firms. Again, the $R^2$ for all cases is small. Panel B indicates that, in the case of the medium and large firms, the significance comes mostly from the period after 1995. In the same period, no factor is important for the first and second quintile. Panel C shows that, in the period before 1996, the current quarter returns and the levels of the current and the previous quarters are important for the smallest firms, and for the fourth quintile.

The above results point to the fact that there is some relationship between the long-term forecasts and the SP 500, but this relationship is not the same for all B/M and size quintiles. The strongest relation appears to be with the growth firms and with the medium and large firms after 1995. Small firms appear to be influenced in the period before 1996. In all cases, the $R^2$ is very small, showing that the SP 500 is not a particularly important factor in explaining the high long-term forecasts. The hypothesis that analysts change their forecasts of long-term growth according to the overall stock market performance is only weakly supported.

**V.B. Firm Returns**

Another possible driving factor for the change in expected growth rates is that analysts become optimistic for firms that achieve high excess returns in the current or the previous quarter; the analysts then extrapolate the market performance of these firms and increase their growth estimates. In the case that the high performance of many firms occurs during the same period (a bull market), their increases in growth estimates could increase the average growth expectations, leading to a synchronized increase in growth estimates. So, the sub-section examines if the movement of forecasts is the result of the returns of certain firms and not so much the result of the overall market behavior. It also provides some preliminary evidence regarding the relation between firm specific returns and long-term forecasts.

I use firm market excess returns over a quarter to examine the impact of firm excess returns on growth expectations. I measure the impact on the growth expectations as the difference between the median expectations (DME) of the current and the previous quarter. As a first step, I present some basic statistics for the cases of positive and negative excess returns in Table [16]. Panel A presents the statistics for the case of positive excess returns. The median of 0% and the mean of -0.133% are too low to offer an explanation for the variation in expectations; moreover, they are negative while the excess return is positive, making it difficult to explain an increase in aggregate expectations. Also, the 25$^{th}$ and 75$^{th}$ centiles are zero, indicating little variance in DME for an individual firm. For the case of negative excess returns, the mean difference is -0.35%, and the median is 0%. The 75$^{th}$ centile is 0%, and the 25$^{th}$ centile is almost 0%. Overall, the results in Table [16] indicate that the firm-specific change in median growth expectations is relatively constant, with most changes equal or close to zero.

I proceed by running a regression with DME as the dependent variable and the excess return of the current quarter as the independent variable. I repeat the regression by replacing the excess return of the current quarter with the excess return of the previous quarter. The results are presented in Table [17] Panel A. The results of the two regressions confirm the results in Table [16] that firm excess returns have little impact on growth expectations. For example, an excess return in the current/previous quarter of 0.01 (1%) will, on average, cause an increase of only 0.0046%/0.0057% in growth expectations. Even if I take into consideration the fact that many DME are zero and run the regressions without them, the beta coefficient of the excess returns increases only to 1.123, implying that an excess return of 1% is associated with 0.01% increase in growth expectations. Another indication is the very low $R^2$, which is virtually zero. A possible objection is that a pooled regression like the one in Panel A conceals the behavior of firms that have certain B/M and size characteristics. Panels B and C of Table [17] present the previous regressions for each B/M and size quintile independently. The results confirm the results of Panel A. The coefficients are too low (even if they are again statistically significant) to indicate any influence of quarter firm excess returns; also, the $R^2$ is again virtually zero for every quintile. Together, the results of Tables [16] and [17] provide evidence that the firm excess returns are, on average, not important for the formation of growth expectations.

Despite the above results, it may be that the influence of the excess returns appears only if the excess return is high or low enough. I choose 10% excess return as the threshold for a high-enough excess return and -10% for a low-enough excess return. I also examine the cases where a firm had at least two consecutive quarters with excess returns of more than 10%. Table [18] Panel A presents the case of a positive firm excess return of 10% or more. The mean change in expected growth rate is -0.13%. For the case of -10% excess return, the mean difference is -0.29%. In both cases, the median is zero. Panels B and C break down the results into B/M and size quintiles. From the three panels, it is evident that the results do not differ greatly from the case where just a positive or negative excess return is required (Table [16] Panels A and B). Requiring persistence of the positive 10% excess return does not alter the conclusions. Table [18] Panel D provides the results for this exercise. In order to have a meaningful change in growth expectation, we need at least six consecutive quarters of 10% excess return. For the case of negative returns (Panel E), it takes three or four quarters for the mean and/or median to achieve a significant size.

In general, the results are not supportive of the hypothesis that analysts increase their expected growth forecasts by taking into consideration the market performance of the firms. The regression coefficients between market returns and forecasted long-term growth are miniscule. Even the case of persistent positive returns cannot adequately explain the phenomenon; those cases are very few and the changes are relatively small compared to the magnitude of returns.

**V.C. GDP Growth**

Another factor that might contribute to the common movement of forecasted growth rates is the GDP growth. The rationale is that the expected growth rates move together because of positive or negative GDP growth, which is a factor that influences the whole economy.

First, I check the correlations between the GDP growth and the long-term expectations for each book-to-market quintile. The results are presented in Table [19]. Panel A shows the results of the B/M quintiles. Only the correlation for the first quintile is significant. However, its economic significance is rather limited, and it is insignificant for the period before 1996. Panel B presents the results for the case of size quintiles, where the correlations for the first quintile are significant for the whole sample and both sub-periods. The correlations for the fifth quintile are negative and significant both economically and statistically, along with the second quintile (but less economically significant here).

As a next step, I run the following regression:

$$D.forecast = a \times LD.forecast + b \times D.GDP\_Growth + c \times LD.GDP\_Growth + \varepsilon \quad (2)$$

where *forecast* is the mean forecast for each quintile. Table [20] Panel A reports the results for the B/M quintiles. For the whole sample, only the coefficient of the difference in GDP growth for the first quintile is significant, indicating that the increase or decrease of GDP growth rate is important for the long-term forecasts of low B/M firms. However, the coefficient is negative, suggesting that an increase in the growth rate is associated with a decrease in long-term forecasts of growth firms. Notice that all coefficients of the change in GDP growth rates are also negative, but insignificant. The results for the low B/M are confirmed in the period after 1995 (Panel B). In contrast, in the period before 1996, no coefficient is significant. The lagged difference of GDP growth is positive for the growth firms and negative for the value firms, but in all cases insignificant and small in size. The $R^2$ is in most cases very low and below 10%, indicating the low ability of the GDP growth rates to explain the changes in long-term forecasts.

Table [21] Panel A reports the results for the size quintiles. For the whole period, only the coefficient of the third quintiles (medium-sized firms) is significant; and it is negative. With the exception of the first quintile, the difference in GDP growth is also negative for all other quintiles (but insignificant). The highest coefficient is of the fifth quintile (largest firms). The situation is similar for the period after 1995; the difference coefficients are negative and the second and third quintiles are significant, indicating a negative relation between change in forecasts and change in GDP growth. For the period before 1996, only the coefficient of the lagged difference of the first quintile is significant, and the remaining coefficients have mixed signs. Similarly to the B/M case, the $R^2$ is very low in most cases.

Overall, the results indicate that change in GDP growth does not adequately explain change in long-term forecasts. There is some explanatory power for the cases of growth firms and medium-sized firms, but the coefficients are negative. There is also some explanatory power for the case of small firms, but it is limited to the first period of the sample. However, GDP growth is not important for the other categories of firms and thus cannot directly explain the synchronized movement of expectations. Moreover, the long-term forecasts are too high to be explained only by GDP growth. The GDP growth over the 30 years of the sample has been relatively low and stable (the Great Moderation period), and it is hard to argue that the analysts were not aware of it. Growing 25% per year for four years while the overall economy grows by around 3% per year during the same period is a difficult task for most firms, even if they are considered to be growth firms.

**V.D. GDP Forecasts**

I turn now to an examination of the relationship between the GDP forecasts and the long-term expectations. The aim is to investigate whether the GDP forecasts may provide an explanation for the high expected growth rates forecasted by analysts. The hypothesis is that the high expected growth rates were based on high expected GDP growth.

The GDP growth forecasts come from the Survey of Professional Forecasters (SPF) conducted every quarter by the Philadelphia FED. I use the mean four-quarters-ahead forecasts. For every quarter, I have a forecast of the average growth for the next 4 quarters, thereby building a time-series of quarterly frequency. Figure [10] depicts the GDP forecasts.

I proceed with an examination of the correlations between the GDP growth forecasts and the long-term growth forecasts for the B/M quintiles. The results are presented at Table [22] Panel A. All correlations are negative while, for the third, fourth, and fifth quintiles, they are significant and economically important. However, if the sample is divided into the periods before 1996 and after 1995, the results change substantially. For the period after 1995, no correlation is significant. For the period before 1996, for the first and third quintiles the correlations are negative and insignificant, for the second quintile positive and significant, and for the last two negative and significant. The results for the case of the size quintiles are reported in Panel B of Table [22]. For the period after 1995, all correlations are insignificant with the exception of quintile four. For the period before 1996, the first and fifth quintile correlations are not significant, the second and fourth positive and significant, and the third negative and significant. Overall, correlations are insignificant after 1995 and do not have homogeneous signs for the period before 1996, so GDP short-term forecasts do not seem to play an important role in shaping long-term forecasts.

I proceed with the following regression:

$$D.forecast = a \times LD.forecast + b \times D.GDP\_Forecast + c \times LD.GDP\_Forecast + \varepsilon \quad (3)$$

The results for the B/M quintiles are in Table [23]. For the whole period, the only significant coefficients are that of the difference in the GDP forecasts at the first B/M quintile (positive) and that of the lagged difference for the fifth quintile (negative). All $R^2$ are very low. The period after 1995 displays a similar picture, with the exception that now it is the lagged difference of the first quintile that is significant. The GDP forecasts are no longer significant for the fifth quintile in the period before 1996. The general picture is that, for the growth firms, an increase in the short-term GDP forecasts means an increase in the long-term forecasts of the growth firms and a decrease in those of the value firms. However the $R^2$ is low and the explanatory ability of the GDP forecasts seems to be low.

Table [24] presents the results for the size quintiles. The only two significant coefficients are that of the lagged difference of the second quintile in the 1996-2011 period and that of the third quintile in the period before 1996. In most cases, the $R^2$ are very small, indicating that the relationship between the difference of GDP forecasts and the difference of long-term forecasts is rather weak.

The above results indicate that the GDP growth forecasts provide some explanation for the growth firms, but not for other firm categories. The above results seem to indicate that factors other than expectations about macroeconomic conditions contribute to the analysts' long-term growth forecasts' excessive optimism about the prospects of growth and large firms. Moreover, the GDP growth forecasts are reasonably accurate; in any case, they do not present an overly rosy prospect for the economy, contrary to the optimistic nature of analysts' long-term forecasts.

**V. E. Profit Forecasts**

The Survey of Professional Forecasters (SPF) by Philadelphia FED includes a questionnaire which asks the participants to forecast the general profit level of all the firms within the economy for the four quarters ahead. The responses could be useful, because they offer another view on the performance of corporations. In the current study, I use operating income and not profits as a performance measure; additionally, the SPR does not include forecasts on individual firms, nor does it distinguish the firms as low or high book-to-market, and as small or large. However, something can be learned if the focus of the analysis is on the growth of profits from quarter to quarter, which can be compared to the growth in the operating performance, which is examined in this study.

For this dataset, the average quarter growth of profits is 1.9% (median 1.5%), with a standard deviation of 1.4%. Figure [11] presents the time series of profit growth forecasts.

Table [25] Panel A presents the correlations between the profit forecasts and analysts' long-term forecasts for the B/M quintiles, as does Panel B for the size quintiles. For the B/M quintiles, all correlations are negative in the period after 1995, as well as the correlations of the third, fourth, and fifth quintiles (the last one not significant) for the period before 1996. For the whole period, all correlations are negative and significant (with the exception of the second quintile). The correlations for the size quintiles are similar. As a preliminary note, negative correlations indicate that financial analysts' forecasts move in opposite directions from the profit forecasts, so profit forecasts by SPF are difficult to use as a factor to explain the movement and values of long-term forecasts. In other words, the expected profits in the economy (according to SPF) cannot be a factor to explain the movement of the long-term forecasts of financial analysts.

The next step is to run the following regression:

$$D.forecast = a \times LD.forecast + b \times D.ProfitForecast + c \times LD.ProfitForecast + \varepsilon \quad (4)$$

The results for the B/M quintiles are in Table [26]. No coefficient is significant for the whole sample. The lagged difference is negative and significant for the value firms in the period after 1995, and the coefficient of the difference is positive and significant for the case of the second quintile in the period before 1996. For the most cases, the $R^2$ is very small.

The results for the size quintiles are in Table [27]. Again, no coefficients are significant for the whole period. There is one negative and significant coefficient for the period after 1995 and two positive and significant coefficients for the period before 1996 for the first and fourth quintiles. In their cases, the $R^2$ is relatively high.

The conclusion is that profit forecasts do not seem to explain the overall value and the movements of analysts' long-term forecasts. In most cases, a change of long-term forecasts does not seem related to the current or previous profit forecasts, and in most cases the $R^2$ is low. Also, financial analysts seem to be excessively optimistic about all the types of firms they cover, in comparison to the forecasters surveyed by the Philadelphia FED. For example, for the first book-to-market quintile (growth) firms, the financial analysts' forecasted growth is consistently above 20% per year for the whole period since 1981. This means an average growth of around 5% per quarter. This is 2.5 times the forecasted profit growth of all firms. For the value firms, the average is around 11% per year, which is around 2.75% quarterly growth, which in turn is 45% more than the average 1.9% forecasted by the SPF. The optimism is exemplified in the forecasts made after 1995. In the case of the size quintiles, the smallest firms (first quintile) have a time series average of 18.86% per year, which in quarterly terms is 4.72%, which is around 2.5 times the average 1.9% forecasted by the SPF. For the largest firms, the average is a 3.125% quarter growth, which is 50% more than the average forecasted by the SPF. So, expectations of future profitability (according to the forecasts of the SPF) do not seem to be the driving factor behind the financial analysts' long-term forecasts.

## V. F. Firm Age

Another factor that could contribute to the co-movement of the growth forecasts is the average firm age, where age is defined as stock market age, that is, the number of quarters that the firm has been publicly traded. It is possible that analysts are excited about the prospects of a new company, while at the same time they discount the prospects of old firms. So, if the mean age of the firms of the B/M and size quintiles declines (for example, because of an IPO wave), the average long-term forecasts could increase. If such a phenomenon exists (that is, a decrease in mean age is associated with an increase in the mean long-term forecast), it could be explained in two ways. The first explanation is that it may be a mechanical effect in the sense that new firms usually come with high long-term forecasts (Michaely and Womack, 1999), so when they enter the stock market the average long-term forecast increases. The other explanation is that analysts may get excited by the number of new incoming firms and increase their expectations about all firms. Notice that not only the firm's age per se is important, but also the meaning of the firm's age. In other words, firm age is used also as a proxy.

Panel A of Table [28] examines the correlations between the forecasted long-term growth for each of the book-to-market quintiles and the average age of the firms of each quintile. For the whole sample, the correlations are negative, significant, and economically important. This is also the case for both sub-periods before 1996 and after 1995. Panel B has the correlations for the size quintiles. Again, the correlations are important for the whole period and also for each sub-period.

I proceed by running the following regression:

$$D.forecast = a \times LD.forecast + b \times D.Age + c \times LD.Age + \varepsilon \quad (5)$$

where *Age* is the mean age of the firms in each quintile.

The results for the B/M case are in Table [29]. For the whole sample, the difference in age is negative and significant for all quintiles. The lagged difference is significant and positive for the fourth quintile, but much smaller than the coefficient of the difference in age, so overall the effect is negative. All $R^2$ are above 10%, and for the first and fourth quintiles are relatively satisfactory. For the period after 1995, the situation is similar, with the exception of the third quintile (which is again negative). The situation is once again the same for the period before 1996, where the $R^2$ for the first quintile is above 50%. Notice also that, in all three cases, the coefficients of the difference in ages are larger (in absolute value) for the growth firms than for the value firms.

The results for the size case are in Table [30]. For the whole sample, the difference in age is negative and significant for all quintiles, as in the B/M case. This is the case also for the period before 1996. For the period after 1995, the difference coefficients is significant for the third, fourth, and fifth quintiles, but not for the first two, even if their coefficients are of the same magnitude as those of the larger quintiles. In many cases, the $R^2$ are satisfactory, and in the case of large firms, above 50%.

Tables [28], [29] and [30] provide evidence that there is an inverse relationship between quintile mean age and quintile mean forecast. This relationship appears to be strong and consistent across periods and across B/M and size quintiles. I examine whether this relationship exists because the new firms entering the stock market cause an increase in expectations. In order to do so, I exclude those firms which have participated in the stock market for three or fewer years, and I calculate again the average expectations. Figures [12], [13] and [14] provide evidence that this is not the case for the B/M quintiles (the second and the fourth quintiles follow the same pattern). There is, of course, a reduction in the long-term expectations, since the new firms come with high growth expectations. However, the difference is small (2-3%); more importantly, the mean forecasted growth still trends in the same way. Notice also that, for the third and fifth quintiles, the older firms have higher expectations and the expectations drift higher. The biggest difference is in the case of the first B/M quintile around 2000, but the mean expectations still drift in the same way. The situation is the same in the case of size quintiles, as depicted in Figures [15], [16] and [17], where the biggest difference is at the first quintile around 2000. (The second and the fourth quintiles follow the same pattern.)

The conclusion is that, when the mean age of the firms decreases, the forecasted long-term growth increases. This effect is independent of the book-to-market and the size of the firms. This fact points to a possible explanation for the high forecasted growth rates and for the co-movement of these expectations for all quintiles: more young firms enter the market; analysts get excited and increase the forecasted growth rates.

## V. G. Operating Income Growth

Another possible factor that could influence analysts in their views about long-term forecasts is the previous year's average growth in operating income. The correlation results are presented in Table [31]. The correlation is positive and significant for the first two quintiles of B/M for all periods. The correlations for the value firms are negative and mainly concentrated in the period before 1996. The correlations are also consistently positive and significant for the last two quintiles in the size classification. They are also positive and significant for the small firms in the period after 1995 and for the firms of the second quintile for the period before 1996. Overall, it seems that the correlations provide evidence that positive average past growth is associated with high forecasted growth for the growth and large firms.

The next step is to run the following regression:

$$D.forecast = a \times LD.forecast + b \times D.OI\_Grwoth + c \times LD.OI\_Growth + \varepsilon \quad (6)$$

where *OI_Growth* is the mean growth of operating income for each B/M and size quintile.

The results for the B/M case are reported in Table [32]. The difference in operating income is significant for the first quintile for all periods. For the second quintile, the coefficients of the difference are significant and positive for the period after 1995 and positive but insignificant for the period before 1996. However, the lagged difference is negative and significant; it seems that the overall effect is negative. The value firms have a negative coefficient for the difference, but its effect is moderated by the positive coefficient of the lagged difference in the whole period regression and the period after 1995. Notice also that the coefficients of the difference for the value firms are relatively small in size compared to those of the growth firms. The $R^2$ for the cases of growth and value firms are moderately satisfactory, but low for the case of intermediate quintiles.

The results for the size quintiles are reported in Table [33]. For the whole period, the coefficients of the difference are positive and significant for the last three quintiles, the medium and large firms. For the period after 1995, the coefficients of the third and fifth quintile remain significant, but the coefficient of the fourth is no longer significant. However its magnitude is at the same level with that of the third quintile. The lagged difference of the small firms is now significant and positive, but it is countered by the equally large negative coefficient of the difference. For the period before 1996, only the coefficient of difference of the big firms is significant. In most cases, the $R^2$ are low.

Overall, there is evidence that past performance is useful in the case of B/M quintiles and of the medium and large firms. In most cases, the coefficients are positive, and the analysts seem to increase their forecasts with an increase in past performance.

**V. H. Interactions with Quarterly Results**

In this subsection, I examine the interaction between quarterly results and changes in the long-term forecasts. Quarterly results that far exceed expectations may cause analysts to increase their long-term forecasts. The same may apply also to results that fall short of expectations. As a measure of expectations, I use the mean forecast of analysts for the given quarter. Because of certain issues with the IBES database, the sample refers to the 1990-2010 period.

I test the above hypothesis by examining the responses of the long-term forecasts to extreme forecast errors, with forecast error defined as the actual value announced after the end of the quarter minus the mean forecast for the given quarter. I define the extreme as the highest 10% and the lowest 10% of the erroneous forecasts. Overly high forecast errors mean that the analysts were too pessimistic (the error here is defined as actual minus forecast) about the firm's quarter results. Overly low forecast errors mean that the analysts were too optimistic about the firm's quarter results. Because missing or exceeding the analysts' mean forecast by a large amount may mean that the firm is taking a deep bath or responding to some other transitory anomaly (perhaps a management change), I also examine the inverse hypothesis: how good or bad are the quarter results in cases of extreme changes in the long-term forecasts.

Table [34] Panel A presents the summary statistics of the forecast error. The 10% and 90% are -0.0667 and 0.09, respectively. I define the change in the long-term forecasts as the change in the next quarter's mean long-term forecast minus the mean long-term forecast of the quarter of the large (in absolute value) forecast error. I also examine a more long-term perspective by calculating the difference in the mean long-term forecasts between that of the same quarter of the following

year and that of the current quarter. For the high forecast errors, the mean difference between that of the current and that of the next quarter is -0.11% (Panel D). For the low (negative) forecast errors, the mean difference is -0.2497% (Panel C). The differences are not economically important; furthermore, they are not significant. In the case of the one-year difference, for the high-forecast errors the mean difference is -0.6367%, while for the low-forecast errors it is -0.86%. Again, they are not significantly different from each other.

I examine the inverse question: how big are the forecast errors in the case of an extreme change in the long-term forecasts? The summary statistics of the change in long-term forecasts are presented in Table [34] Panel B. For the case of an excessive (above 90%) increase in the long-term forecasts in the period of one quarter, the mean forecast error is -0.0008615. For the case of an excessive decline (below 10%), the mean forecast error is 0.0017785. The difference between the two is not statistically significant.

As a result, it seems that the short-term forecast error is not a crucial factor in determining the long-term forecasts. In other words, analysts do not become optimistic or pessimistic because of extraordinary short-term results. Also, when analysts increase (or decrease) their long-term forecasts, they are not influenced by the short-term results of the covered firm.

## V. I. Conclusion and Further Investigation

The analysis of the previous subsection leads to certain conclusions. First, firm performance is not the most important factor in analysts' forecasts, at least as measured by firm excess returns and earnings surprises. Second, certain macro factors may explain some part of overall forecasting behavior, but not all of it; for example, GDP is related to certain quintiles but not to all. This may indicate that macro factors have some influence on the formation of expectations. The third conclusion is that long-term forecasts are mostly influenced by the mean age of the firms in each quintile and by the previous mean performance of the quintile. The fourth conclusion is that security analysts are often much more optimistic than their counterparts who focus on the whole economy.

As a next step, I run a regression including both the difference in mean age and the difference in previous performance, and their lagged values. The results for the case of B/M quintiles are presented in Table [35]. The results for the whole period indicate that difference in mean age is negative and significant for all quintiles, with the coefficient of the first quintile being the largest (in absolute value). The difference in previous performance is positive and significant for the first three quintiles, insignificant for the fourth, and negative and insignificant for the fifth. The results are similar for the two sub-periods, with the notable exception that the difference in past performance is negative and significant for the value firms. In most cases now, the $R^2$ is relatively satisfactory. Including the other factors studied previously increases the $R^2$, but most additional variables are not significant.

The results for the size quintiles are in Table [36]. Difference in quintile age is negative and significant for all quintiles, while difference in past performance is significant only for the last three quintiles. For the period after 1995, there is no significant factor for the first two quintiles, while for the period before 1996, only the difference in age is significant. The conclusion from Tables [35] and [36] is that both differences in age and past performance are related to the difference in long-term forecasts; this relationship is stronger for growth firms and large firms. This provides evidence that the Extrapolation Hypothesis can be associated more with overestimation of the growth of growth and large firms, than with underestimation of the growth of value and small firms. As in the B/M case, including the other factors studied previously increases the $R^2$, but most additional variables are not significant.

A possible explanation for the relationship of the long-term estimates with mean age and past performance is that of technological advancement. The first channel through which technological advancement could increase the long-term forecasts is by improving the current performance and also the prospects of future performance. This may lead analysts to overestimate the future prospects of firms. The second channel is by creating opportunities for new firms, which then go public, seeking access to capital. The entrance of new firms may create euphoria among analysts, who then raise their forecasts. Such an explanation implies that increases in long-term forecasts will be higher in the periods of above-average IPO activity and TFP growth (a proxy for technological advancement), that differences in mean age will be lower (the mean age decreases) in periods of above-average IPO activity, and that differences in growth of operating income will be higher in periods of higher TFP growth. These effects should be more prominent for growth and large firms.

In order to conduct a preliminary investigation of the above hypothesis, I divide the sample into periods of above- and below-median IPO[2] and above- and below-median TFP[3] growth. I then examine the difference in growth of mean forecasts, mean age, and mean operating income growth for each period. The results for the B/M case are in Table [37]. Regarding IPOs, the results are positive for all quintiles with respect to forecasts (indicating higher forecast growth in a hot IPO period) and negative (except for the value firms) for mean age (mean age declines more in a hot IPO period). The results are significant for the long-term forecasts and quintile age in the case of growth firms. Regarding TFP, the results are significant for the long-term forecasts in the case of the growth firms (indicating higher growth of forecasts in periods of high TFP growth), and for all firms in terms of operating income. Notice that the difference in forecast is negative for value firms, meaning that analysts decrease their forecasts during high TFP growth. The results for the size quintiles are in Table [38]. Regarding IPOs, the results are significant for the long-term forecasts and quintile age in the case of the large firms, and are also positive for the first two quintiles. Regarding TFP, the results are significant for the long-term forecasts in the case of the large firms, and for all firms in terms of operating income. Overall, the results in Tables [37] and [38] provide some very preliminary evidence that the movement in analysts' forecasts may be triggered by the number of IPOs and the growth of TFP, especially in the case of growth and large firms.

---

[2] The IPO dataset is kindly provided by Jay Ritter at http://bear.warrington.ufl.edu/ritter/

[3] The TFP dataset is kindly provided by John Fernald at http://www.frbsf.org/economic-research/total-factor-productivity-tfp/

# VI. <u>Conclusions</u>

I have examined the performance of long-term analysts' growth forecasts. Analysts tended to overestimate the growth potential of most firms for most of the period from 1981 to 2011. The overestimation is prevalent among growth and large firms, while the overestimation in the case of value and small firms is much more moderate, allowing for periods of underestimation. The errors associated with growth firms are always higher than those associated with value firms; and the errors are also higher for large firms than for small firms, providing support for the Extrapolation Hypothesis. Another aspect of analysts' long-term forecasts is that they move together for all firm categories, even if they stay at different levels. I identify two common factors behind the over-optimism and co-movement of expectations: the mean age of firms and the previous mean performance. On the other hand, forecasts do not seem to be much influenced by overall market returns, firm excess returns, current GDP growth, GDP growth forecasts, overall corporate profit forecasts, and earnings surprises. The two factors that influence long-term expectations provide some support for the idea that analysts' optimism could be triggered by some productivity-enhancing technological advancement, which causes the analysts to become more optimistic regarding the future prospects of firms.

Overall, the results point to the facts that analysts are overly optimistic and that they do not take into consideration the mean reversion inherent in the performance of firms in competitive markets (Fama and French, 1995). If analysts' forecasts are to be taken as a proxy for investors' expectations and forecasts, then investors are likewise prone to naïve extrapolation.

**Table 1**. Expectations and expectation errors for the four-year horizon for each book-to-market quintile

| Book-to-market Quintile | Simple Error | Relative Error | Long-Term Growth Forecast |
|---|---|---|---|
| 1 | 5.824758 | 0.215747 | 20 |
| 2 | 3.920853 | 0.145143 | 15 |
| 3 | 2.126846 | 0.069286 | 12.5 |
| 4 | 1.434441 | 0.037311 | 11 |
| 5 | 1.021879 | 0.012685 | 10.68 |

Table 1: This table reports the median simple error, the median relative error, and the median growth forecast for each book-to-market quintile. The forecasts were issued by analysts for the period 1981-2011 and are long-term forecasts, estimating average annual growth of operating income for the following three to five years. I use four years as an average between three and five years. The simple error is defined as the four-years-ahead forecasted operating income minus the actual operating income four years later. The relative error is defined as the simple error but is divided by the actual operating income four years later. The long-term growth forecasts are the forecasted annual growth rates in percent. The accounting data come from Compustat. Stock data come from CRSP. The long-term forecasts come from IBES.

**Table 2**. Book-to-market quintile expectations and expectation errors by exchange for the four-year horizon

| Book-to-market Quintile | | | NYSE | AmEx | NASDAQ |
|---|---|---|---|---|---|
| | 1 | Relative Error | 0.23213 | 0.15545 | 0.2016 |
| | | Growth forecast (%) | 15 | 21.125 | 25 |
| | 2 | Relative Error | 0.18167 | 0.19714 | 0.08673 |
| | | Growth forecast (%) | 13 | 15 | 17.5 |
| | 3 | Relative Error | 0.09923 | 0.02513 | 0.02648 |
| | | Growth forecast (%) | 11 | 13 | 15 |
| | 4 | Relative Error | 0.05195 | 0.08112 | -0.0157 |
| | | Growth forecast (%) | 9.5 | 15 | 14.5 |
| | 5 | Relative Error | 0.02406 | 0.1838 | -0.0859 |
| | | Growth forecast (%) | 8.695 | 14.15 | 15 |

Table 2: This table reports the median relative error and the median growth forecast for each book-to-market quintile and each stock market. The forecasts were issued by analysts for the period 1981-2011 and are long-term forecasts, estimating average annual growth of operating income for the following three to five years. I use four years as an average between three and five years. The simple error is defined as the four-years-ahead forecasted operating income minus the actual operating income four years later. The relative error is defined as the simple error but is divided by the actual operating income four years later. The long-term growth forecasts are the forecasted annual growth rates in percent. The accounting data come from Compustat. Stock data come from CRSP. The long-term forecasts come from IBES.

**Table 3.** Book-to-market quintile expectations and expectation errors for the two-year horizon

| Book-to-market Quintile | Simple Error | Relative Error | Long-Term Growth Forecast |
|---|---|---|---|
| 1 | 1.40394 | 0.06655 | 20 |
| 2 | 1.188509 | 0.050976 | 15 |
| 3 | 0.8018498 | 0.031007 | 12.5 |
| 4 | 0.4819975 | 0.012345 | 11 |
| 5 | 0.3237357 | 0.004301 | 10.68 |

Table 3: This table reports the median simple error, the median relative error, and the median growth forecast for each book-to-market quintile. The forecasts were issued by analysts for the period 1981-2011 and are long-term forecasts, estimating an average annual growth of operating income for the following three to five years. The simple error is defined as the two-years-ahead forecasted operating income minus the actual operating income two years later. The relative error is defined as the simple error but is divided by the actual operating income two years later. The long-term growth forecasts are the forecasted annual growth rates in percent. The accounting data come from Compustat. Stock data come from CRSP. The long-term forecasts come from IBES.

**Table 4.** Book-to-market quintile expectations and expectation errors by exchange for the two-year horizon

| Book-to-market decile | | | NYSE | AmEx | NASDAQ |
|---|---|---|---|---|---|
| | 1 | Relative Error | 0.08762 | 0.06742 | 0.04595 |
| | | Growth forecast (%) | 15 | 21.125 | 25 |
| | 2 | Relative Error | 0.07829 | 0.08884 | 0.01801 |
| | | Growth forecast (%) | 13 | 15 | 17.5 |
| | 3 | Relative Error | 0.05663 | 0.03714 | -0.0013 |
| | | Growth forecast (%) | 11 | 13 | 15 |
| | 4 | Relative Error | 0.02033 | 0.02918 | -0.0123 |
| | | Growth forecast (%) | 9.5 | 15 | 14.5 |
| | 5 | Relative Error | 0.00957 | 0.01734 | -0.0199 |
| | | Growth forecast (%) | 8.695 | 14.15 | 15 |

Table 4: This table reports the median relative error, and the median growth forecast for each book-to-market quintile by stock market. The forecasts were issued by analysts for the period 1981-2011 and are long-term forecasts, estimating an average annual growth of operating income for the following three to five years. The simple error is defined as the two-years-ahead forecasted operating income minus the actual operating income two years later. The relative error is defined as the simple error but is divided by the actual operating income two years later. The long-term growth forecasts are the forecasted annual growth rates in percent. The accounting data come from Compustat. Stock data come from CRSP. The long-term forecasts come from IBES.

**Table 5.** Size quintile expectations and expectation errors for the four-year horizon

| Size Quintile | Simple Error | Relative Error | Long-Term Growth Forecast |
|---|---|---|---|
| 1 | 0.394376 | 0.006722 | 16.5 |
| 2 | 2.020374 | 0.075275 | 15 |
| 3 | 4.683392 | 0.117076 | 14 |
| 4 | 12.9744 | 0.154531 | 12 |
| 5 | 45.19731 | 0.130472 | 11 |

Table 5: This table reports the median simple error, the median relative error, and the median growth forecast for each size quintile. The forecasts were issued by analysts for the period 1981-2011 and are long-term forecasts, estimating an average annual growth of operating income for the following three to five years. I use four years as an average between three and five years. The simple error is defined as the four-years-ahead forecasted operating income minus the actual operating income four years later. The relative error is defined as the simple error but is divided by the actual operating income four years later. The long-term growth forecasts are the forecasted annual growth rates in percent. The accounting data come from Compustat. Stock data come from CRSP. The long-term forecasts come from IBES.

**Table 6.** Size quintile expectations and expectation errors by exchange for the four-year horizon

| Size Quintile | | NYSE | AmEx | NASDAQ |
|---|---|---|---|---|
| 1 | Relative Error | 0.02886 | 0.04704 | -0.0004 |
|   | Growth forecast (%) | 12.9 | 15 | 18 |
| 2 | Relative Error | 0.04163 | 0.09823 | 0.10361 |
|   | Growth forecast (%) | 13 | 15 | 18 |
| 3 | Relative Error | 0.11305 | 0.14144 | 0.12095 |
|   | Growth forecast (%) | 12 | 15 | 17 |
| 4 | Relative Error | 0.14854 | 0.24193 | 0.16124 |
|   | Growth forecast (%) | 11 | 15 | 17.5 |
| 5 | Relative Error | 0.12125 | 0.87553 | 0.18767 |
|   | Growth forecast (%) | 10 | 14.125 | 18 |

Table 6: The table reports the median relative error and the median growth forecast for each size quintile and each stock market. The forecasts were issued by analysts for the period 1981-2011 and are long-term forecasts, estimating an average annual growth of operating income for the following three to five years. I use four years as an average between three and five years. The simple error is defined as the four-years-ahead forecasted operating income minus the actual operating income four years later. The relative error is defined as the simple error but is divided by the actual operating income four years later. The long-term growth forecasts are the forecasted annual growth rates in percent. The accounting data come from Compustat. Stock data come from CRSP. The long-term forecasts come from IBES.

**Table 7.** Size quintile expectations and expectation errors for the two-year horizon

| Size Quintile | Simple Error | Relative Error | Long-Term Growth Forecast |
|---|---|---|---|
| 1 | 0.082824 | -0.008 | 16.5 |
| 2 | 0.604317 | 0.019336 | 15 |
| 3 | 1.781837 | 0.047455 | 14 |
| 4 | 4.71273 | 0.05938 | 12 |
| 5 | 17.17621 | 0.053816 | 11 |

Table 7: This table reports the median simple error, the median relative error, and the median growth forecast for each size quintile. The forecasts were issued by analysts for the period 1981-2011 and are long-term forecasts, estimating an average annual growth of operating income for the following three to five years. The simple error is defined as the two-years-ahead forecasted operating income minus the actual operating income two years later. The relative error is defined as the simple error but is divided by the actual operating income two years later. The long-term growth forecasts are the forecasted annual growth rates in percent. The accounting data come from Compustat. Stock data come from CRSP. The long-term forecasts come from IBES.

**Table 8**. Size quintile expectations and expectation errors by exchange for the two-year horizon

| Size Quintile | | NYSE | AmEx | NASDAQ |
|---|---|---|---|---|
| 1 | Relative Error | 0.02135 | 0.01366 | -0.0148 |
|   | Growth forecast | 12.9 | 15 | 18 |
| 2 | Relative Error | 0.01525 | 0.04845 | 0.02219 |
|   | Growth forecast | 13 | 15 | 18 |
| 3 | Relative Error | 0.05191 | 0.0538 | 0.03826 |
|   | Growth forecast | 12 | 15 | 17 |
| 4 | Relative Error | 0.06339 | 0.11011 | 0.03639 |
|   | Growth forecast | 11 | 15 | 17.5 |
| 5 | Relative Error | 0.05329 | 0.2632 | 0.05246 |
|   | Growth forecast | 10 | 14.125 | 18 |

Table 8: This table reports the median relative error, and the median growth forecast for each size quintile. The forecasts were issued by analysts for the period 1981-2011 and are long-term forecasts, estimating an average annual growth of operating income for the following three to five years. The simple error is defined as the two-years-ahead forecasted operating income minus the actual operating income two years later. The relative error is defined as the simple error but is divided by the actual operating income two years later. The long-term growth forecasts are the forecasted annual growth rates in percent. The accounting data come from Compustat. Stock data come from CRSP. The long-term forecasts come from IBES.

**Table 9.** Size and book-to-market quintile relative errors for the four-year horizon

|  | Book-to-market quintile | | | | |
|---|---|---|---|---|---|
|  | 1 | 2 | 3 | 4 | 5 |
| size quintile | | | | | |
| 1 | 0.1073413 | 0.0345812 | 0.0360601 | -0.02088 | -0.0808132 |
| 2 | 0.2253785 | 0.1015829 | 0.0289075 | 0.0155379 | 0.0089975 |
| 3 | 0.2334118 | 0.1870988 | 0.0349756 | 0.0470354 | 0.0493386 |
| 4 | 0.2495524 | 0.1860047 | 0.1574714 | 0.0683645 | 0.0526462 |
| 5 | 0.210604 | 0.165409 | 0.1174748 | 0.060166 | 0.0135154 |

Table 9: This table reports the median relative error for each book-to-market and size quintile. The forecasts were issued by analysts for the period 1981-2011 and are long-term forecasts, estimating an average annual growth of operating income for the following three to five years. I use four years as an average between three and five years. The simple error is defined as the four-years-ahead forecasted operating income minus the actual operating income four years later. The relative error is defined as the simple error but is divided by the actual operating income four years later. The long-term growth forecasts are the forecasted annual growth rates in percent. The accounting data come from Compustat. Stock data come from CRSP. The long-term forecasts come from IBES.

**Table 10**. EPS forecasts with respect to previous FYE actual EPS results (% growth)

Panel A: B/M quintiles

| B/M quintile | Median (%) increase of forecasts relative to the previous actual results | Mean (%) increase of forecasts relative to the previous actual results |
|---|---|---|
| 1 | 0.1636364 | 0.1639388 |
| 2 | 0.1312779 | 0.1376213 |
| 3 | 0.1009174 | 0.0976733 |
| 4 | 0.0740741 | 0.0819844 |
| 5 | 0.0311284 | -0.0039447 |

Panel B: Size quintiles

| Size quintile | Median (%) increase of forecasts relative to the previous actual results | Mean (%) increase of forecasts relative to the previous actual results |
|---|---|---|
| 1 | 0.1135135 | 0.064406 |
| 2 | 0.1386139 | 0.1076823 |
| 3 | 0.1386139 | 0.1192375 |
| 4 | 0.1290323 | 0.1263238 |
| 5 | 0.1235955 | 0.1318214 |

Table 10: This table reports the growth of EPS forecasts at the beginning of the year, relative to the actual EPS results for the previous year for each B/M (Panel A) and size (Panel B) quintile for the period 1981-2011. The forecasts were issued by analysts in the period 1981-2011 and are EPS forecasts for the current FYE. The accounting data come from Compustat. Stock data come from CRSP. The EPS forecasts come from IBES.

**Table 11.** Long-term growth forecast and error correlations among book-to-market quintiles

Panel A: Long term growth forecast correlations

| B/M quintile | 1 | 2 | 3 | 4 | 5 |
|---|---|---|---|---|---|
| 1 | 1.00 | | | | |
| 2 | 0.79*** | 1.00 | | | |
| 3 | 0.74*** | 0.76*** | 1.00 | | |
| 4 | 0.60*** | 0.65*** | 0.90*** | 1.00 | |
| 5 | 0.59*** | 0.57*** | 0.76*** | 0.87*** | 1.00 |

Panel B: Relative error correlations

| B/M quintile | 1 | 2 | 3 | 4 | 5 |
|---|---|---|---|---|---|
| 1 | 1.00 | | | | |
| 2 | 0.70*** | 1.00 | | | |
| 3 | 0.81*** | 0.73*** | 1.00 | | |
| 4 | 0.74*** | 0.53*** | 0.73*** | 1.00 | |
| 5 | 0.72*** | 0.54*** | 0.65*** | 0.80*** | 1.00 |

Table 11: This table reports the correlations between each two book-to-market quintiles for the case of long-term forecasts (Panel A) and for the case of relative errors (Panel B) for each book-to-market quintile for the period 1981-2011. The forecasts were issued by analysts for the period 1981-2011 and are long-term forecasts, estimating an average annual growth of operating income for the following three to five years. I use four years as an average between three and five years. The relative error is defined as the simple error (forecast minus actual) but is divided by the actual operating income four years later. The long-term growth forecasts are the forecasted annual growth rates in percent. The accounting data come from Compustat. Stock data come from CRSP. The long-term forecasts come from IBES. One star indicates significance at the 10%, two stars at the 5% and three at the 1% level.

**Table 12.** Long-term growth forecast and error correlations among size quintiles

Panel A: Long-term growth forecast correlations for size quintiles

| Size quintile | 1 | 2 | 3 | 4 | 5 |
|---|---|---|---|---|---|
| 1 | 1.00 | | | | |
| 2 | 0.83*** | 1.00 | | | |
| 3 | 0.81*** | 0.94*** | 1.00 | | |
| 4 | 0.67*** | 0.83*** | 0.90*** | 1.00 | |
| 5 | 0.59*** | 0.78*** | 0.88*** | 0.96*** | 1.00 |

Panel B: Relative error correlations

| Size quintile | 1 | 2 | 3 | 4 | 5 |
|---|---|---|---|---|---|
| 1 | 1.00 | | | | |
| 2 | 0.77*** | 1.00 | | | |
| 3 | 0.73*** | 0.88*** | 1.00 | | |
| 4 | 0.67*** | 0.81*** | 0.87*** | 1.00 | |
| 5 | 0.72*** | 0.77*** | 0.84*** | 0.85*** | 1.00 |

Table 12: This table reports the correlations between each two size quintiles for the case of long-term forecasts (Panel A) and for the case of relative errors (Panel B) for each size quintile for the period 1981-2011. The forecasts were issued by analysts for the period 1981-2011 and are long-term forecasts, estimating an average annual growth of operating income for the following three to five years. I use four years as an average between three and five years. The relative error is defined as the simple error (forecast minus actual) but is divided by the actual operating income four years later. The long-term growth forecasts are the forecasted annual growth rates in percent. The accounting data come from Compustat. Stock data come from CRSP. The long-term forecasts come from IBES. One star indicates significance at the 10%, two stars at the 5% and three at the 1% level.

**Table 13**. Correlation of SP 500 and long-term forecasts

Panel A: Pearson correlations for B/M quintiles

| | whole sample | =>1996 | <=1995 |
|---|---|---|---|
| 1 | 0.34*** | 0.08 | 0.50*** |
| 2 | 0.36*** | -0.03 | -0.02 |
| 3 | 0.65*** | 0.06 | 0.39*** |
| 4 | 0.81*** | 0.03 | 0.81*** |
| 5 | 0.67*** | 0.08 | 0.80*** |

Panel B: Spearman correlations for B/M quintiles

| | whole sample | =>1996 | <=1995 |
|---|---|---|---|
| 1 | 0.25*** | 0.09 | 0.44*** |
| 2 | 0.39*** | -0.01 | -0.06 |
| 3 | 0.69*** | 0.11 | 0.29** |
| 4 | 0.82*** | -0.02 | 0.77*** |
| 5 | 0.80*** | 0.16 | 0.82*** |

Panel C: Pearson correlations for size quintiles

| | whole sample | =>1996 | <=1995 |
|---|---|---|---|
| 1 | 0.34*** | -0.02 | 0.39*** |
| 2 | 0.49*** | -0.01 | 0.65*** |
| 3 | 0.58*** | 0.14 | 0.83*** |
| 4 | 0.66*** | 0.28** | 0.42*** |
| 5 | 0.67*** | 0.38*** | 0.18 |

Panel D: Spearman correlations for size quintiles

|   | whole sample | =>1996 | <=1995 |
|---|---|---|---|
| 1 | 0.34*** | -0.09 | 0.44*** |
| 2 | 0.50*** | -0.01 | 0.59*** |
| 3 | 0.66*** | 0.06 | 0.85*** |
| 4 | 0.71*** | 0.12 | 0.28** |
| 5 | 0.78*** | 0.44*** | 0.31** |

Table 13: This table reports the correlations between the long-term growth forecasts of each book-to-market (Panels A and B) and size (Panels C and D) quintiles with the SP 500. In all Panels the data come with quarterly frequency. The forecasts were issued by analysts for the period 1981-2011; they are long-term forecasts, estimating an average annual growth of operating income for the following three to five years. The long-term growth forecasts are the forecasted annual growth rates in percent. Market returns data come from CRSP. The accounting data come from Compustat. The long-term forecasts come from IBES. One star indicates significance at the 10%, two stars at the 5% and three at the 1% level.

**Table 14**. Influence of SP 500 on the long-term forecasts of the B/M quintiles

Panel A: Whole sample

|  | 1st B/M | 2nd B/M | 3rd B/M | 4th B/M | 5th B/M |
|---|---|---|---|---|---|
|  | Change in quintile aggregate estimates | | | | |
| Previous quarter Change of estimates | 0.06 | -0.16* | 0.02 | -0.05 | -0.08 |
| Current quarter return of SP 500 | -0.38 | 4.41* | -1.08 | 0.16 | 3.11 |
| Previous quarter return of SP 500 | 3.21** | 1.29 | -0.14 | 0.02 | -0.50 |
| Current quarter level of SP 500 | 0.00 | -0.01* | 0.00 | -0.00 | -0.00 |
| Previous quarter level of SP 500 | -0.00 | 0.01** | -0.00 | 0.00 | 0.00 |
| Constant | -0.14 | -0.19 | 0.07 | 0.10 | -0.08 |
| N | 117 | 117 | 117 | 117 | 117 |
| R-sq | 0.074 | 0.067 | 0.005 | 0.041 | 0.017 |

\* $p<0.10$     \*\* $p<0.05$     \*\*\* $p<0.01$

Panel B: 1996-2011

|  | 1st B/M | 2nd B/M | 3rd B/M | 4th B/M | 5th B/M |
|---|---|---|---|---|---|
|  | Change in quintile aggregate estimates | | | | |
| Previous quarter Change of estimates | 0.06 | -0.00 | 0.02 | 0.01 | -0.08 |
| Current quarter return of SP 500 | -0.47 | 8.33 | 4.67 | -4.78 | 5.50 |
| Previous quarter return of SP 500 | 5.03** | 2.86* | 0.20 | 0.98 | -0.82 |
| Current quarter level of SP 500 | 0.00 | -0.01 | -0.00 | 0.00 | -0.01 |
| Previous quarter level of SP 500 | -0.00 | 0.01 | 0.00 | -0.00 | 0.01 |
| Constant | -0.97 | -0.69 | -0.35 | 0.56 | -0.70 |
| N | 62 | 62 | 62 | 62 | 62 |
| R-sq | 0.142 | 0.101 | 0.009 | 0.061 | 0.017 |

* $p<0.10$ \qquad ** $p<0.05$ \qquad *** $p<0.01$

Panel C: 1981-1995

|  | 1st B/M | 2nd B/M | 3rd B/M | 4th B/M | 5th B/M |
|---|---|---|---|---|---|
|  | Change in quintile aggregate estimates | | | | |
| Previous quarter Change of estimates | -0.10 | -0.30** | -0.07 | -0.22 | -0.06 |
| Current quarter return of SP 500 | 1.25 | 17.25*** | -2.34 | -2.50 | 1.20 |
| Previous quarter return of SP 500 | -0.20 | -1.26 | -0.84 | -1.06 | -0.18 |
| Current quarter level of SP 500 | -0.00 | -0.05** | 0.00 | 0.01 | 0.00 |
| Previous quarter level of SP 500 | 0.01 | 0.05** | -0.00 | -0.01 | -0.00 |
| Constant | -0.15 | -0.97** | -0.05 | 0.06 | 0.04 |
| N | 55 | 55 | 55 | 55 | 55 |
| R-sq | 0.019 | 0.212 | 0.049 | 0.115 | 0.070 |

* $p<0.10$ \qquad ** $p<0.05$ \qquad *** $p<0.01$

Table 14: This table reports the regression results using the SP 500 as the independent variable. Panel A reports the results for the whole 1981-2011 period, Panel B for the 1996-2011 period, and Panel C for the 1981-1995 period. In all Panels, the data come with quarterly frequency. The forecasts were issued by analysts for the period 1981-2011; they are long-term forecasts, estimating an average annual growth of operating income for the following three to five years. The long-term growth forecasts are the forecasted annual growth rates in percent. The accounting data come from Compustat. Market returns data come from CRSP. The long-term forecasts come from IBES.

**Table 15.** Influence of SP 500 on the long-term forecasts of the size quintiles

Panel A: Whole sample

|  | 1st Size | 2nd Size | 3rd Size | 4th Size | 5th Size |
|---|---|---|---|---|---|
|  | Change in quintile aggregate estimates | | | | |
| Previous quarter Change of estimates | -0.19** | -0.03 | 0.05 | 0.05 | -0.00 |
| Current quarter return of SP 500 | 6.15** | -0.03 | 0.58 | -1.90 | -1.77 |
| Previous quarter return of SP 500 | -0.33 | 1.19 | 2.52*** | 2.28** | 2.29** |
| Current quarter level of SP 500 | -0.01** | 0.00 | -0.00 | 0.00 | 0.00 |
| Previous quarter level of SP 500 | 0.01** | -0.00 | 0.00 | -0.00 | -0.00 |
| Constant | -0.04 | 0.01 | -0.10 | -0.13 | -0.20 |
| N | 117 | 117 | 117 | 117 | 117 |
| R-sq | 0.075 | 0.016 | 0.070 | 0.079 | 0.093 |

* $p<0.10$    ** $p<0.05$    *** $p<0.01$

Panel B: 1996-2011

|  | 1st Size | 2nd Size | 3rd Size | 4th Size | 5th Size |
|---|---|---|---|---|---|
|  | Change in quintile aggregate estimates | | | | |
| Previous quarter Change of estimates | -0.02 | 0.01 | 0.06 | 0.01 | -0.05 |
| Current quarter return of SP 500 | 0.65 | 9.99 | 1.07 | 1.39 | -7.04 |
| Previous quarter return of SP 500 | 1.16 | 2.01 | 3.87*** | 4.10** | 3.86** |
| Current quarter level of SP 500 | -0.00 | -0.01 | -0.00 | 0.00 | 0.01 |
| Previous quarter level of SP 500 | 0.00 | 0.01 | 0.00 | -0.00 | -0.01 |
| Constant | -0.07 | -1.11 | -0.64 | -1.10 | -0.94 |
| N | 62 | 62 | 62 | 62 | 62 |
| R-sq | 0.031 | 0.071 | 0.133 | 0.136 | 0.183 |

* $p<0.10$    ** $p<0.05$    *** $p<0.01$

Panel C: 1981-1995

|  | 1st Size | 2nd Size | 3rd Size | 4th Size | 5th Size |
|---|---|---|---|---|---|
|  | Change in quintile aggregate estimates | | | | |
| Previous quarter Change of estimates | -0.36*** | -0.32** | -0.10 | -0.20 | -0.07 |
| Current quarter return of SP 500 | 19.52*** | -2.87 | 2.48 | -5.13** | -0.50 |
| Previous quarter return of SP 500 | -2.03 | -0.38 | 0.08 | -0.96 | -0.72 |
| Current quarter level of SP 500 | -0.05** | 0.01 | -0.01 | 0.02** | 0.00 |
| Previous quarter level of SP 500 | 0.06** | -0.01 | 0.01 | -0.02** | 0.00 |
| Constant | -0.48 | 0.47 | -0.11 | 0.00 | -0.25 |
| N | 55 | 55 | 55 | 55 | 55 |
| R-sq | 0.245 | 0.120 | 0.031 | 0.230 | 0.117 |

* $p<0.10$       ** $p<0.05$       *** $p<0.01$

Table 15: This table reports the regression results using the SP 500 as the independent variable. Panel A reports the results for the whole 1981-2011 period, Panel B for the 1996-2011 period, and Panel C for the 1981-1995 period. In all Panels, the data come with quarterly frequency. The forecasts were issued by analysts for the period 1981-2011; they are long-term forecasts, estimating an average annual growth of operating income for the following three to five years. The long-term growth forecasts are the forecasted annual growth rates in percent. The accounting data come from Compustat. Market returns data come from CRSP. The long-term forecasts come from IBES.

**Table 16**. Influence of positive and negative excess firm returns on the long-term forecasts

Panel A: Positive excess returns

| Percentile | 10 | 25 | 50 | 75 | 90 | Mean | St. Dev. |
|---|---|---|---|---|---|---|---|
| Quarter change in estimates | -2 | 0 | 0 | 0 | 1.5 | -0.133 | 4.405 |

Panel B: Negative excess returns

| Percentile | 10 | 25 | 50 | 75 | 90 | Mean | St. Dev. |
|---|---|---|---|---|---|---|---|
| Quarter change in estimates | -2.5 | -0.1 | 0 | 0 | 1 | -0.35 | 4.179 |

Table 16: This table reports the change in median long-term forecasts for the case of positive (Panel A) and negative (Panel B) market excess returns of the current quarter. In all Panels, the data come with quarterly frequency. The forecasts were issued by analysts for the period 1981-2011; they are long-term forecasts, estimating an average annual growth of operating income for the following three to five years. The long-term growth forecasts are the forecasted annual growth rates in percent. The accounting data come from Compustat. Market returns data come from CRSP. The long-term forecasts come from IBES.

**Table 17**. Regression estimates using the firm excess returns as independent variable

Panel A: Whole sample

|  | whole sample | whole sample |
|---|---|---|
|  | DME | |
| Quarter Excess Return | 0.46*** | |
|  | (15.01) | |
| Previous Quarter Excess Return |  | 0.57*** |
|  |  | (19.36) |
| Constant | -0.21*** | -0.22*** |
|  | (-39.14) | (-39.69) |
| N | 134270 | 127376 |
| R-sq | 0.002 | 0.006 |

t statistics in parentheses
* $p<0.10$   ** $p<0.05$   *** $p<0.001$

Panel B: B/M quintiles

| | 1st B/M | 1st B/M | 2nd B/M | 2nd B/M | 3rd B/M | 3rd B/M | 4th B/M | 4th B/M | 5th B/M | 5th B/M |
|---|---|---|---|---|---|---|---|---|---|---|
| | DME | | DME | | DME | | DME | | DME | |
| Quarter Excess Return | 0.63*** | | 0.54*** | | 0.32*** | | 0.33*** | | 0.33*** | |
| Previous Quarter Excess Return | | 0.78*** | | 0.60*** | | 0.46*** | | 0.49*** | | 0.41*** |
| Constant | -0.27*** | -0.30*** | -0.20*** | -0.21*** | -0.16*** | -0.16*** | -0.13*** | -0.14*** | -0.20*** | -0.20*** |
| N | 33823 | 31415 | 28313 | 26961 | 27340 | 26158 | 25184 | 24092 | 20316 | 19399 |
| R-sq | 0.004 | 0.009 | 0.003 | 0.006 | 0.002 | 0.004 | 0.002 | 0.005 | 0.002 | 0.004 |

\* $p<0.10$    \*\* $p<0.05$    \*\*\* $p<0.001$

Panel C: Size quintiles

|  | 1st Size | 1st Size | 2nd Size | 2nd Size | 3rd Size | 3rd Size | 4th Size | 4th Size | 5th Size | 5th Size |
|---|---|---|---|---|---|---|---|---|---|---|
|  | DME | DME | DME | DME | DME | DME | DME | DME | DME | DME |
| Quarter Excess Return | 0.50*** |  | 0.52*** |  | 0.36*** |  | 0.36*** |  | 0.36*** |  |
| Previous Quarter Excess Return |  | 0.54*** |  | 0.63*** |  | 0.65*** |  | 0.55*** |  | 0.57*** |
| Constant | -0.27*** |  | -0.26*** |  | -0.15*** |  | -0.10*** |  | -0.11*** |  |
|  |  | -0.29*** |  | -0.28*** |  | -0.16*** |  | -0.11*** |  | -0.12*** |
| N | 33823 | 31415 | 28313 | 26961 | 27340 | 26158 | 25184 | 24092 | 20316 | 19399 |
| R-sq | 0.004 | 0.009 | 0.003 | 0.006 | 0.002 | 0.004 | 0.002 | 0.005 | 0.002 | 0.004 |

*p<0.10      ** p<0.05      *** p<0.001

Table 17: This table reports the regression results using the firm excess returns as the independent variable and the difference in median expectations (DME) as the dependent variable. Panel A reports the results for the whole sample, Panel B for each B/M quintile separately, and Panel C for each size quintile. In all Panels, the data come with quarterly frequency. The forecasts were issued by analysts for the period 1981-2011; they are long-term forecasts, estimating an average annual growth of operating income for the following three to five years. The long-term growth forecasts are the forecasted annual growth rates in percent. The accounting data come from Compustat. Market returns data come from CRSP. The long-term forecasts come from IBES.

**Table 18**. Difference in median expectation (DME) and large excess returns

Panel A: Whole sample

|  | mean(DME) | med(DME) | N(DME) |
|---|---|---|---|
| Less than -10 % excess return | -0.2957739 | 0 | 99,084 |
| More than 10% excess return | -0.1286853 | 0 | 40,716 |

Panel B: B/M qunitiles

|  |  | 1.00 | 2.00 | 3.00 | 4.00 | 5.00 |
|---|---|---|---|---|---|---|
| Less than -10 % excess return | Mean | -0.43 | -0.29 | -0.22 | -0.22 | -0.28 |
|  | Median | 0.00 | 0.00 | 0.00 | 0.00 | 0.00 |
| More than 10% excess return | Mean | -0.18 | -0.13 | -0.11 | -0.09 | -0.10 |
|  | Median | 0.00 | 0.00 | 0.00 | 0.00 | 0.00 |

Panel C: Size quintiles

|  |  | 1.00 | 2.00 | 3.00 | 4.00 | 5.00 |
|---|---|---|---|---|---|---|
| Less than -10 % excess return | Mean | -0.41 | -0.38 | -0.26 | -0.15 | -0.15 |
|  | Median | 0.00 | 0.00 | 0.00 | 0.00 | 0.00 |
| More than 10% excess return | Mean | -0.21 | -0.15 | -0.09 | -0.03 | -0.03 |
|  | Median | 0.00 | 0.00 | 0.00 | 0.00 | 0.00 |

Panel D: Consecutive quarters of equal to or more than 10% firm excess return

| # of quarters in a row | mean(cum. DME) | median (cum. DME) |
|---|---|---|
| 1 | -0.2510205 | 0 |
| 2 | -0.1911248 | 0 |
| 3 | 0.1584099 | 0 |
| 4 | 1.170526 | 0 |
| 5 | 0.327707 | 0 |
| 6 | 2.18875 | 1.65 |
| 7 | -0.4261905 | 2 |
| 8 | 3.833333 | 3 |
| 9 | 10.75 | 10.75 |

Panel E: Consecutive quarters of equal to or less than -10% firm excess return

| # of quarters in a row | mean(cum. DME) | median (cum. DME) |
|---|---|---|
| 1 | -0.28336 | 0 |
| 2 | -1.06701 | 0 |
| 3 | -3.02058 | -0.55 |
| 4 | -3.1728 | -1.25 |
| 5 | -6.81946 | -3.5 |
| 6 | -10.3044 | -7.25 |
| 7 | -8.75 | -7.25 |
| 8 | -4.83333 | -6 |
| 10 | -20 | -20 |
| 11 | -20 | -20 |

Table 18: This table reports the DME with respect to large firm excess returns. Panel A reports the results for an excess return equal to or more (less) than 10% (-10%) for the whole sample, Panel B reports the same results for each B/M quintile, and Panel C for each size quintile. Panel D examines the results of DME regarding consecutive quarters of +10% or more excess return and Panel E the results of DME regarding consecutive quarters of -10% or less excess return. In all Panels, the data come with quarterly frequency. The forecasts were issued by analysts for the period 1981-2011; they are long-term forecasts, estimating an average annual growth of operating income for the following three to five years. The long-term growth forecasts are the forecasted annual growth rates in percent. The accounting data come from Compustat. Market returns data come from CRSP. The long-term forecasts come from IBES.

**Table 19**. Correlation of GDP growth rates and long-term forecasts

Panel A: B/M quintiles

|   | whole sample | >1995 | <=1995 |
|---|---|---|---|
| 1 | 0.15** | 0.24* | 0.19 |
| 2 | 0.13 | 0.19 | 0.21 |
| 3 | 0.00 | 0.14 | -0.01 |
| 4 | -0.07 | 0.08 | -0.06 |
| 5 | -0.12 | -0.06 | -0.10 |

Panel B: Size quintiles

|   | whole sample | >1995 | <=1995 |
|---|---|---|---|
| 1 | 0.24*** | 0.22* | 0.51*** |
| 2 | 0.08 | 0.14 | 0.22* |
| 3 | 0.030 | 0.13 | 0.07 |
| 4 | -0.016 | 0.09 | -0.04 |
| 5 | -0.14 | -0.04 | -0.35*** |

Table 19: This table reports the Pearson correlations of the long-term growth forecasts of each book-to-market (Panel A) and size (Panel B) quintile with the GDP growth rates. In all Panels, the data come with quarterly frequency. The forecasts were issued by analysts for the period 1981-2011; they are long-term forecasts, estimating an average annual growth of operating income for the following three to five years. The long-term growth forecasts are the forecasted annual growth rates in percent. Market returns data come from CRSP. The accounting data come from Compustat. The long-term forecasts come from IBES. One star means significance at the 10%, two at the 5%, and three at the 1% level.

**Table 20**. Influence of GDP growth rates on the long-term forecasts of the B/M quintiles

Panel A: Whole Sample

|  | 1st B/M | 2nd B/M | 3rd B/M | 4th B/M | 5th B/M |
|---|---|---|---|---|---|
|  | Change in quintile aggregate estimates | | | | |
| Previous quarter Change of estimates | 0.16** | -0.13 | 0.02 | -0.07 | -0.07 |
| D.GDP growth rate | -42.30* | -5.37 | -9.52 | -12.14 | -23.60 |
| LD.GDP growth rate | 20.26 | 24.75 | 3.64 | -10.43 | -10.12 |
| Constant | -0.02 | -0.01 | 0.00 | 0.03 | 0.05 |
| N | 117 | 117 | 117 | 117 | 117 |
| R-sq | 0.103 | 0.057 | 0.011 | 0.016 | 0.019 |

* p<0.10      ** p<0.05    *** p<0.01

Panel B: 1996-2011

|  | 1st B/M | 2nd B/M | 3rd B/M | 4th B/M | 5th B/M |
|---|---|---|---|---|---|
|  | Change in quintile aggregate estimates | | | | |
| Previous quarter Change of estimates | 0.21** | -0.00 | 0.03 | -0.04 | -0.07 |
| D.GDP growth rate | -72.86* | -31.59 | -6.26 | -23.53 | -43.18 |
| LD.GDP growth rate | 17.26 | 1.46 | 4.10 | -24.65 | -14.71 |
| Constant | -0.06 | -0.04 | -0.03 | -0.00 | 0.04 |
| N | 62 | 62 | 62 | 62 | 62 |
| R-sq | 0.163 | 0.052 | 0.006 | 0.032 | 0.033 |

* p<0.10      ** p<0.05    *** p<0.01

Panel C: 1981-1995

|  | 1st B/M | 2nd B/M | 3rd B/M | 4th B/M | 5th B/M |
|---|---|---|---|---|---|
|  | Change in quintile aggregate estimates | | | | |
| Previous quarter Change of estimates | -0.06 | -0.27 | -0.03 | -0.20** | -0.05 |
| D.GDP growth rate | -1.44 | 25.66 | -15.04 | -0.91 | 0.95 |
| LD.GDP growth rate | 24.80 | 57.26 | 1.61 | 6.16 | -5.29 |
| Constant | 0.03 | 0.00 | 0.05 | 0.06 | 0.04 |
| N | 55 | 55 | 55 | 55 | 55 |
| R-sq | 0.072 | 0.173 | 0.029 | 0.054 | 0.007 |

\* $p<0.10$    \*\* $p<0.05$    \*\*\* $p<0.01$

Table 20: This table reports the regression results using the difference in GDP growth rates as the independent variable and the difference in mean forecasted long-term growth rates of each B/M quintile as the dependent variable. Panel A reports the results for the whole 1981-2011 period, Panel B for the 1996-2011 period, and Panel C for the 1981-1995 period. In all Panels, the data come with quarterly frequency. The forecasts were issued by analysts for the period 1981-2011; they are long-term forecasts, estimating an average annual growth of operating income for the next three to five years. The long-term growth forecasts are the forecasted annual growth rates in percent. The accounting data come from Compustat. Market returns data come from CRSP. The long-term forecasts come from IBES.

**Table 21**. Influence of GDP growth rates on the long-term forecasts of the size quintiles

Panel A: Whole Sample

|  | 1st Size | 2nd Size | 3rd Size | 4th Size | 5th Size |
|---|---|---|---|---|---|
|  | Change in quintile aggregate estimates | | | | |
| Previous quarter Change of estimates | -0.16* | 0.03 | 0.11 | 0.01 | 0.07 |
| D.GDP growth rate | 4.12 | -28.62 | -32.44* | -22.81 | -33.62 |
| LD.GDP growth rate | 42.15 | -3.22 | -8.03 | 3.75 | -3.10 |
| Constant | 0.02 | 0.02 | 0.01 | 0.01 | -0.01 |
| N | 117 | 117 | 117 | 117 | 117 |
| R-sq | 0.079 | 0.047 | 0.067 | 0.023 | 0.072 |

\* $p<0.10$    \*\* $p<0.05$    \*\*\* $p<0.01$

Panel B: 1996-2011

|  | 1st Size | 2nd Size | 3rd Size | 4th Size | 5th Size |
|---|---|---|---|---|---|
|  | Change in quintile aggregate estimates | | | | |
| Previous quarter Change of estimates | -0.06 | 0.15 | 0.23*** | 0.16*** | 0.08 |
| D.GDP growth rate | -23.66 | -47.45* | -68.05** | -54.92 | -54.20 |
| LD.GDP growth rate | 2.40 | -5.04 | -5.99 | 3.65 | -9.96 |
| Constant | -0.03 | -0.02 | -0.03 | 0.01 | 0.01 |
| N | 62 | 62 | 62 | 62 | 62 |
| R-sq | 0.024 | 0.096 | 0.206 | 0.095 | 0.107 |

\* p<0.10      \*\* p<0.05      \*\*\* p<0.01

Panel C: 1981-1995

|  | 1st Size | 2nd Size | 3rd Size | 4th Size | 5th Size |
|---|---|---|---|---|---|
|  | Change in quintile aggregate estimates | | | | |
| Previous quarter Change of estimates | -0.30** | -0.29*** | -0.20 | -0.47** | 0.05 |
| D.GDP growth rate | 36.98 | -6.52 | 12.79 | 18.15 | -7.13 |
| LD.GDP growth rate | 93.93** | -0.69 | -0.93 | 16.90 | 5.59 |
| Constant | 0.06 | 0.10 | 0.04 | -0.01 | -0.04 |
| N | 55 | 55 | 55 | 55 | 55 |
| R-sq | 0.276 | 0.096 | 0.077 | 0.246 | 0.041 |

\* p<0.10      \*\* p<0.05      \*\*\* p<0.01

Table 21: This table reports the regression results using the difference in GDP growth rates as the independent variable and the difference in mean forecasted long-term growth rates of each size quintile as the dependent variable. Panel A reports the results for the whole 1981-2011 period, Panel B for the 1996-2011 period, and Panel C for the 1981-1995 period. In all Panels, the data come with quarterly frequency. The forecasts were issued by analysts for the period 1981-2011; they are long-term forecasts, estimating an average annual growth of operating income for the following three to five years. The long-term growth forecasts are the forecasted annual growth rates in percent. The accounting data come from Compustat. Market returns data come from CRSP. The long-term forecasts come from IBES.

**Table 22**. Correlation of forecasts of GDP growth rates and long-term forecasts

Panel A: B/M quintiles

|  | whole sample | >1995 | <=1995 |
|---|---|---|---|
| 1 | -0.15 | 0.17 | -0.17 |
| 2 | -0.11 | 0.12 | 0.37*** |
| 3 | -0.44*** | 0.06 | -0.16 |
| 4 | -0.61*** | -0.00 | -0.50*** |
| 5 | -0.57*** | -0.06 | -0.69*** |

Panel B: Size quintiles

|   | whole sample | >1995 | <=1995 |
|---|---|---|---|
| 1 | -0.20** | 0.16 | -0.17 |
| 2 | -0.38*** | 0.10 | 0.52*** |
| 3 | -0.41*** | 0.07 | -0.66*** |
| 4 | -0.25*** | 0.22* | 0.25* |
| 5 | -0.29*** | 0.14 | 0.19 |

Table 22: This table reports the correlations of the long-term growth forecasts of each book-to-market (Panel A) and size (Panel B) quintiles with the forecasts of the GDP growth rates. In all Panels, the data come with quarterly frequency. The forecasts were issued by analysts for the period 1981-2011; they are long-term forecasts, estimating an average annual growth of operating income for the following three to five years. The long-term growth forecasts are the forecasted annual growth rates in percent. Market returns data come from CRSP. The accounting data come from Compustat. The long-term forecasts come from IBES. GDP forecasts data come from the Philadelphia FED. One star indicates significance at the 10%, two stars at the 5% and three at the 1% level.

**Table 23**. Influence of forecasts of GDP growth rates on the long-term forecasts of the B/M quintiles

Panel A: Whole Sample

|   | 1st B/M | 2nd B/M | 3rd B/M | 4th B/M | 5th B/M |
|---|---|---|---|---|---|
|   | Change in quintile aggregate estimates | | | | |
| Previous quarter Change of estimates | 0.06 | -0.16 | 0.02 | -0.08 | -0.10 |
| L.GDP forecast (4 quarters) | 95.56* | 52.79 | -36.00 | -82.00 | -89.34 |
| LD.GDP forecast (4 quarters) | 95.96 | 32.05 | 8.08 | -35.45 | -132.67** |
| Constant | -0.01 | -0.00 | -0.00 | 0.02 | 0.03 |
| N | 117 | 117 | 117 | 117 | 117 |
| R-sq | 0.042 | 0.034 | 0.006 | 0.036 | 0.045 |

\* p<0.10    \*\* p<0.05    \*\*\* p<0.01

Panel B: 1996-2011

|   | 1st B/M | 2nd B/M | 3rd B/M | 4th B/M | 5th B/M |
|---|---|---|---|---|---|
|   | Change in quintile aggregate estimates | | | | |
| Previous quarter Change of estimates | 0.09 | -0.05 | 0.03 | -0.07 | -0.11 |
| L.GDP forecast (4 quarters) | 64.33 | -16.27 | -38.00 | -132.36 | -183.30 |
| LD.GDP forecast (4 quarters) | 156.71* | 111.37 | 0.61 | -48.65 | -164.47* |
| Constant | -0.07 | -0.04 | -0.03 | 0.00 | 0.04 |
| N | 62 | 62 | 62 | 62 | 62 |
| R-sq | 0.055 | 0.037 | 0.007 | 0.057 | 0.072 |

\* p<0.10    \*\* p<0.05    \*\*\* p<0.01

Panel C: 1981-1995

|  | 1st B/M | 2nd B/M | 3rd B/M | 4th B/M | 5th B/M |
|---|---|---|---|---|---|
|  | Change in quintile aggregate estimates | | | | |
| Previous quarter Change of estimates | -0.08 | -0.25 | -0.02 | -0.21** | -0.02 |
| L.GDP forecast (4 quarters) | 127.12** | 155.49 | -27.17 | 9.43 | 69.04 |
| LD.GDP forecast (4 quarters) | 25.76 | -58.68 | 22.84 | 3.24 | -66.96 |
| Constant | 0.06 | 0.05 | 0.04 | 0.06 | 0.04 |
| N | 55 | 55 | 55 | 55 | 55 |
| R-sq | 0.085 | 0.121 | 0.008 | 0.046 | 0.062 |

\* $p<0.10$ \qquad ** $p<0.05$ \qquad *** $p<0.01$

Table 23: This table reports the regression results using the difference in forecasts of GDP growth rates as the independent variable and the difference in mean forecasted long-term growth rates of each B/M quintile as the dependent variable. Panel A reports the results for the whole 1981-2011 period, Panel B for the 1996-2011 period, and Panel C for the 1981-1995 period. In all Panels, the data come with quarterly frequency. The forecasts were issued by analysts for the period 1981-2011; they are long-term forecasts, estimating an average annual growth of operating income for the following three to five years. The long-term growth forecasts are the forecasted annual growth rates in percent. The accounting data come from Compustat. Market returns data come from CRSP. The long-term forecasts come from IBES. GDP forecasts data come from the Philadelphia FED.

**Table 24**. Influence of forecasts of GDP growth rates on the long-term forecasts of the size quintiles

Panel A: Whole Sample

|  | 1st Size | 2nd Size | 3rd Size | 4th Size | 5th Size |
|---|---|---|---|---|---|
|  | Change in quintile aggregate estimates | | | | |
| Previous quarter Change of estimates | -0.18** | -0.02 | 0.06 | -0.02 | 0.04 |
| D.GDP forecast (4 quarters) | 28.81 | -7.29 | -3.65 | 67.49 | 26.92 |
| LD.GDP forecast (4 quarters) | -34.46 | 43.50 | 45.91 | 56.24 | 18.23 |
| Constant | 0.03 | 0.02 | 0.00 | 0.01 | -0.01 |
| N | 117 | 117 | 117 | 117 | 117 |
| R-sq | 0.035 | 0.007 | 0.012 | 0.017 | 0.006 |

\* $p<0.10$ \qquad ** $p<0.05$ \qquad *** $p<0.01$

Panel B: 1996-2011

|  | 1st Size | 2nd Size | 3rd Size | 4th Size | 5th Size |
|---|---|---|---|---|---|
|  | Change in quintile aggregate estimates | | | | |
| Previous quarter Change of estimates | -0.07 | 0.07 | 0.13 | 0.09 | 0.03 |
| D.GDP forecast (4 quarters) | -78.06 | -38.03 | -64.14 | 49.14 | 24.81 |
| LD.GDP forecast (4 quarters) | 23.29 | 89.70* | 106.21 | 103.22 | 58.72 |
| Constant | -0.03 | -0.02 | -0.02 | 0.01 | 0.01 |
| N | 62 | 62 | 62 | 62 | 62 |
| R-sq | 0.016 | 0.028 | 0.051 | 0.033 | 0.012 |

\* $p<0.10$     \*\* $p<0.05$    \*\*\* $p<0.01$

Panel C: 1981-1995

|  | 1st Size | 2nd Size | 3rd Size | 4th Size | 5th Size |
|---|---|---|---|---|---|
|  | Change in quintile aggregate estimates | | | | |
| Previous quarter Change of estimates | -0.27** | -0.32*** | -0.18 | -0.47** | 0.05 |
| D.GDP forecast (4 quarters) | 198.17 | 44.25 | 81.88** | 83.01 | 16.81 |
| LD.GDP forecast (4 quarters) | -60.97 | -7.61 | -4.49 | 12.99 | -49.09 |
| Constant | 0.13 | 0.10 | 0.06 | 0.02 | -0.04 |
| N | 55 | 55 | 55 | 55 | 55 |
| R-sq | 0.142 | 0.103 | 0.108 | 0.242 | 0.047 |

\* $p<0.10$     \*\* $p<0.05$    \*\*\* $p<0.01$

Table 24: This table reports the regression results using the difference in forecasts of GDP growth rates as the independent variable and the difference in mean forecasted long-term growth rate of each size quintile as the dependent variable. Panel A reports the results for the whole 1981-2011 period, Panel B for the 1996-2011 period, and Panel C for the 1981-1995 period. In all Panels, the data come with quarterly frequency. The forecasts were issued by analysts for the period 1981-2011; they are long-term forecasts, estimating an average annual growth of operating income for the following three to five years. The long-term growth forecasts are the forecasted annual growth rates in percent. The accounting data come from Compustat. Market returns data come from CRSP. The long-term forecasts come from IBES. GDP forecasts data come from the Philadelphia FED.

**Table 25**. Correlation of corporate profit forecasts and long-term forecasts

Panel A: B/M quintiles

|  | whole sample | >1995 | <=1995 |
|---|---|---|---|
| 1 | -0.27*** | -0.39*** | 0.02 |
| 2 | -0.11 | -0.41*** | 0.44*** |
| 3 | -0.39*** | -0.35*** | -0.39*** |
| 4 | -0.28*** | -0.16 | -0.27** |
| 5 | -0.20** | -0.08 | -0.11 |

Panel B: Size quintiles

|   | whole sample | >1995 | <=1995 |
|---|---|---|---|
| 1 | -0.12 | -0.23* | 0.20 |
| 2 | -0.31*** | -0.29** | -0.28** |
| 3 | -0.35*** | -0.38*** | -0.30** |
| 4 | -0.19** | -0.30** | 0.41*** |
| 5 | -0.21** | -0.29** | 0.30** |

Table 25: This table reports the correlations of the long-term growth forecasts of each book-to-market (Panel A) and size (Panel B) quintiles with the forecasts of corporate profits. In all Panels, the data come with quarterly frequency. The forecasts were issued by analysts for the period 1981-2011; they are long-term forecasts estimating an average annual growth of operating income for the following three to five years. The long-term growth forecasts are the forecasted annual growth rates in percent. Market returns data come from CRSP. The accounting data come from Compustat. The long-term forecasts come from IBES. Corporate profit forecasts data come from the Philadelphia FED. One star indicates significance at the 10%, two stars at the 5% and three at the 1% level.

**Table 26**. Influence of forecasts of corporate profits on the long-term forecasts of the B/M quintiles

Panel A: Whole Sample

|  | 1st B/M | 2nd B/M | 3rd B/M | 4th B/M | 5th B/M |
|---|---|---|---|---|---|
|  | Change in quintile aggregate estimates | | | | |
| Previous quarter Change of estimates | 0.07 | -0.17 | -0.01 | -0.09 | -0.08 |
| D.Corporate profits forecast (4 quarters) | -5.00 | 15.76 | -8.00 | -5.27 | -2.21 |
| LD.Corporate profits forecast (4 quarters) | 14.59 | 12.89 | -5.81 | -8.17 | -6.07 |
| Constant | -0.02 | -0.01 | 0.00 | 0.03 | 0.04 |
| N | 117 | 117 | 117 | 117 | 117 |
| R-sq | 0.019 | 0.052 | 0.012 | 0.016 | 0.008 |

\* p<0.10    \*\* p<0.05    \*\*\* p<0.01

Panel B: 1996-2011

|  | 1st B/M | 2nd B/M | 3rd B/M | 4th B/M | 5th B/M |
|---|---|---|---|---|---|
|  | Change in quintile aggregate estimates | | | | |
| Previous quarter Change of estimates | 0.12 | -0.05 | 0.01 | -0.08 | -0.11 |
| D.Corporate profits forecast (4 quarters) | -19.50 | -20.46 | -8.24 | -14.49 | -29.34 |
| LD.Corporate profits forecast (4 quarters) | 38.05 | 17.10 | -5.84 | -9.69 | -39.02* |
| Constant | -0.06 | -0.04 | -0.03 | 0.01 | 0.05 |
| N | 62 | 62 | 62 | 62 | 62 |
| R-sq | 0.071 | 0.051 | 0.007 | 0.015 | 0.041 |

\* $p<0.10$   \*\* $p<0.05$   \*\*\* $p<0.01$

Panel C: 1981-1995

|  | 1st B/M | 2nd B/M | 3rd B/M | 4th B/M | 5th B/M |
|---|---|---|---|---|---|
|  | Change in quintile aggregate estimates | | | | |
| Previous quarter Change of estimates | -0.07 | -0.35* | -0.06 | -0.21** | -0.10 |
| D.Corporate profits forecast (4 quarters) | 11.13 | 47.14** | -7.96 | 1.21 | 8.92 |
| LD.Corporate profits forecast (4 quarters) | -2.82 | 10.10 | -6.17 | -9.58 | 10.74 |
| Constant | 0.04 | 0.05 | 0.04 | 0.06 | 0.05 |
| N | 55 | 55 | 55 | 55 | 55 |
| R-sq | 0.035 | 0.252 | 0.026 | 0.088 | 0.060 |

\* $p<0.10$   \*\* $p<0.05$   \*\*\* $p<0.01$

Table 26: This table reports the regression results using the difference in forecasts of corporate profits as the independent variable and the difference in mean forecasted long-term growth rate of each B/M quintile as the dependent variable. Panel A reports the results for the whole 1981-2011 period, Panel B for the 1996-2011 period, and Panel C for the 1981-1995 period. In all Panels, the data come with quarterly frequency. The forecasts were issued by analysts for the period 1981-2011; they are long-term forecasts, estimating an average annual growth of operating income for the following three to five years. The long-term growth forecasts are the forecasted annual growth rates in percents. The accounting data come from Compustat. Market returns data come from CRSP. The long-term forecasts come from IBES. Corporate profits forecasts data come from the Philadelphia FED.

**Table 27**. Influence of forecasts of corporate profits on the long-term forecasts of the size quintiles

Panel A: Whole Sample

|  | 1st Size | 2nd Size | 3rd Size | 4th Size | 5th Size |
|---|---|---|---|---|---|
|  | Change in quintile aggregate estimates | | | | |
| Previous quarter Change of estimates | -0.17* | -0.01 | 0.07 | -0.01 | 0.04 |
| D.Corporate profits forecast (4 quarters) | 25.23 | -7.17 | -9.67 | -0.57 | -7.70 |
| LD.Corporate profits forecast (4 quarters) | 16.87 | 6.98 | 8.11 | 7.37 | 2.28 |
| Constant | 0.03 | 0.02 | -0.00 | 0.00 | -0.01 |
| N | 117 | 117 | 117 | 117 | 117 |
| R-sq | 0.072 | 0.011 | 0.023 | 0.003 | 0.008 |

* $p<0.10$  ** $p<0.05$  *** $p<0.01$

Panel B: 1996-2011

|  | 1st Size | 2nd Size | 3rd Size | 4th Size | 5th Size |
|---|---|---|---|---|---|
|  | Change in quintile aggregate estimates | | | | |
| Previous quarter Change of estimates | -0.09 | 0.09 | 0.12 | 0.11** | 0.03 |
| D.Corporate profits forecast (4 quarters) | -14.68 | -15.46 | -26.36* | -22.69 | -20.59 |
| LD.Corporate profits forecast (4 quarters) | -2.10 | 24.31 | 13.57 | 19.23 | 6.03 |
| Constant | -0.03 | -0.02 | -0.02 | 0.01 | 0.01 |
| N | 62 | 62 | 62 | 62 | 62 |
| R-sq | 0.011 | 0.058 | 0.070 | 0.044 | 0.026 |

* $p<0.10$  ** $p<0.05$  *** $p<0.01$

Panel C: 1981-1995

|  | 1st Size | 2nd Size | 3rd Size | 4th Size | 5th Size |
|---|---|---|---|---|---|
|  | Change in quintile aggregate estimates | | | | |
| Previous quarter Change of estimates | -0.36*** | -0.30*** | -0.20 | -0.55*** | 0.04 |
| D.Corporate profits forecast (4 quarters) | 48.81* | 3.04 | 5.05 | 25.84* | 2.67 |
| LD.Corporate profits forecast (4 quarters) | 31.61 | -5.22 | 2.98 | 5.33 | -3.76 |
| Constant | 0.13 | 0.09 | 0.05 | 0.01 | -0.04 |
| N | 55 | 55 | 55 | 55 | 55 |
| R-sq | 0.285 | 0.100 | 0.057 | 0.340 | 0.015 |

* $p<0.10$  ** $p<0.05$  *** $p<0.01$

Table 27: This table reports the regression results using the difference in forecasts of corporate profits as the independent variable and the difference in mean forecasted long-term growth rates of each size quintile as the dependent variable. Panel A reports the results for the whole 1981-2011 period, Panel B for the 1996-2011 period, and Panel C for the 1981-1995 period. In all Panels, the data come with quarterly frequency. The forecasts were issued by analysts for the period 1981-2011; they are long-term forecasts estimating an average annual growth of operating income for the following three to five years. The long-term growth forecasts are the forecasted annual growth rates in percent. The accounting data come from Compustat. Market returns data come from CRSP. The long-term forecasts come from IBES. Corporate profits forecasts data come from the Philadelphia FED.

**Table 28**. Correlation of mean quintile firm age with long-term forecasts

Panel A: B/M quintiles

|   | whole sample | >1995 | <=1995 |
|---|---|---|---|
| 1 | -0.76*** | -0.89*** | -0.81*** |
| 2 | -0.63*** | -0.73*** | -0.29** |
| 3 | -0.64*** | -0.69*** | -0.35*** |
| 4 | -0.61*** | -0.71*** | -0.63*** |
| 5 | -0.70*** | -0.66*** | -0.70*** |

Panel B: Size quintiles

|   | whole sample | >1995 | <=1995 |
|---|---|---|---|
| 1 | -0.51*** | -0.72*** | -0.33** |
| 2 | -0.60*** | -0.76*** | -0.52*** |
| 3 | -0.77*** | -0.80*** | -0.64*** |
| 4 | -0.80*** | -0.79*** | -0.39*** |
| 5 | -0.71*** | -0.81*** | -0.26* |

Table 28: This table reports the correlations between the long-term growth forecasts of each book-to-market (Panel A) and size (Panel B) quintile with the mean quintile firm age. Firm age is defined as the number of quarters the firm is publicly traded. In all Panels, the data come with quarterly frequency. The forecasts were issued by analysts for the period 1981-2011; they are long-term forecasts, estimating an average annual growth of operating income for the following three to five years. The long-term growth forecasts are the forecasted annual growth rates in percent. Market returns data come from CRSP. The accounting data come from Compustat. The long-term forecasts come from IBES. One star indicates significance at the 10%, two stars at the 5% and three at the 1% level.

**Table 29**. Influence of mean quintile firm age on the long-term forecasts of the B/M quintiles

Panel A: Whole Sample

|  | 1st B/M | 2nd B/M | 3rd B/M | 4th B/M | 5th B/M |
|---|---|---|---|---|---|
|  | Change in quintile aggregate estimates ||||| 
| Previous quarter Change of estimates | -0.02 | -0.17 | 0.03 | 0.04 | -0.06 |
| Change in quintile mean firm age | -0.28*** | -0.15*** | -0.06* | -0.12*** | -0.11*** |
| Previous change in quintile mean firm age | -0.06 | -0.03 | 0.01 | 0.04*** | 0.02 |
| Constant | 0.04 | 0.02 | 0.02 | 0.04 | 0.02 |
| N | 117 | 117 | 117 | 117 | 117 |
| R-sq | 0.354 | 0.161 | 0.114 | 0.331 | 0.144 |

\* p<0.10     \*\* p<0.05     \*\*\* p<0.01

Panel B: 1996-2011

|  | 1st B/M | 2nd B/M | 3rd B/M | 4th B/M | 5th B/M |
|---|---|---|---|---|---|
|  | Change in quintile aggregate estimates ||||| 
| Previous quarter Change of estimates | -0.01 | -0.01 | 0.03 | 0.07 | -0.07 |
| Change in quintile mean firm age | -0.29** | -0.16* | -0.05 | -0.14** | -0.16** |
| Previous change in quintile mean firm age | -0.05 | 0.03 | 0.00 | 0.07*** | 0.04 |
| Constant | 0.12 | 0.03 | 0.00 | 0.06 | 0.03 |
| N | 62 | 62 | 62 | 62 | 62 |
| R-sq | 0.324 | 0.181 | 0.097 | 0.367 | 0.172 |

\* p<0.10     \*\* p<0.05     \*\*\* p<0.01

Panel C: 1981-1995

|  | 1st B/M | 2nd B/M | 3rd B/M | 4th B/M | 5th B/M |
|---|---|---|---|---|---|
|  | Change in quintile aggregate estimates ||||| 
| Previous quarter Change of estimates | -0.15 | -0.31 | 0.08 | -0.05 | 0.07 |
| Change in quintile mean firm age | -0.27*** | -0.15*** | -0.08** | -0.08*** | -0.06** |
| Previous change in quintile mean firm age | -0.12** | -0.10 | 0.03** | 0.01 | 0.01 |
| Constant | -0.04 | -0.03 | 0.04 | 0.03 | 0.02 |
| N | 55 | 55 | 55 | 55 | 55 |
| R-sq | 0.544 | 0.200 | 0.183 | 0.405 | 0.203 |

\* p<0.10     \*\* p<0.05     \*\*\* p<0.01

Table 29: This table reports the regression results using the difference in mean quintile firm age as the independent variable and the difference in mean forecasted long-term growth rates of each B/M quintile as the dependent variable. Firm age is defined as the number of quarters the firm is publicly traded. Panel A reports the results for the whole 1981-2011 period, Panel B for the 1996-2011 period, and Panel C for the 1981-1995 period. In all Panels, the data come with quarterly frequency. The forecasts were issued by analysts for the period 1981-2011; they are long-term forecasts estimating an average annual growth of operating income for the following three to five years. The long-term growth forecasts are the forecasted annual growth rates in percent. The accounting data come from Compustat. Market returns data come from CRSP. The long-term forecasts come from IBES.

**Table 30**. Influence of forecasts of mean quintile firm age on the long-term forecasts of the size quintiles

Panel A: Whole Sample

|  | 1st Size | 2nd Size | 3rd Size | 4th size | 5th Size |
|---|---|---|---|---|---|
|  | Change in quintile aggregate estimates | | | | |
| Previous quarter Change of estimates | -0.15 | 0.02 | 0.10 | -0.01 | 0.09 |
| Change in quintile mean firm age | -0.12** | -0.07** | -0.11*** | -0.11** | -0.13*** |
| Previous change in quintile mean firm age | 0.04 | 0.01 | -0.00 | -0.02 | 0.01 |
| Constant | 0.03 | 0.02 | 0.02 | 0.03 | 0.01 |
| N | 117 | 117 | 117 | 117 | 117 |
| R-sq | 0.138 | 0.080 | 0.189 | 0.289 | 0.529 |

\* $p<0.10$   \*\* $p<0.05$   \*\*\* $p<0.01$

Panel B: 1996-2011

|  | 1st Size | 2nd Size | 3rd Size | 4th size | 5th Size |
|---|---|---|---|---|---|
|  | Change in quintile aggregate estimates | | | | |
| Previous quarter Change of estimates | -0.08 | 0.05 | 0.19* | 0.22*** | 0.09 |
| Change in quintile mean firm age | -0.16 | -0.11 | -0.20** | -0.17*** | -0.14*** |
| Previous change in quintile mean firm age | 0.01 | 0.00 | 0.04 | 0.02 | 0.01 |
| Constant | 0.04 | 0.03 | 0.07 | 0.06 | -0.00 |
| N | 62 | 62 | 62 | 62 | 62 |
| R-sq | 0.141 | 0.069 | 0.313 | 0.533 | 0.579 |

\* $p<0.10$   \*\* $p<0.05$   \*\*\* $p<0.01$

Panel C: 1981-1995

|  | 1st Size | 2nd Size | 3rd Size | 4th size | 5th Size |
|---|---|---|---|---|---|
|  | Change in quintile aggregate estimates | | | | |
| Previous quarter Change of estimates | -0.24* | -0.20** | -0.15 | -0.46** | 0.06 |
| Change in quintile mean firm age | -0.07* | -0.05* | -0.04* | -0.02* | -0.06*** |
| Previous change in quintile mean firm age | 0.07 | -0.00 | -0.02* | -0.01* | 0.00 |
| Constant | 0.10 | 0.06 | 0.03 | 0.01 | -0.01 |
| N | 55 | 55 | 55 | 55 | 55 |
| R-sq | 0.179 | 0.174 | 0.173 | 0.233 | 0.253 |

* $p<0.10$     ** $p<0.05$     *** $p<0.01$

Table 30: This table reports the regression results using the difference in mean quintile age as the independent variable and the difference in mean forecasted long-term growth rates of each size quintile as the dependent variable. Firm age is defined as the number of quarters the firm is publicly traded. Panel A reports the results for the whole 1981-2011 period, Panel B for the 1996-2011 period, and Panel C for the 1981-1995 period. In all Panels, the data come with quarterly frequency. The forecasts were issued by analysts for the period 1981-2011; they are long-term forecasts estimating an average annual growth of operating income for the following three to five years. The long-term growth forecasts are the forecasted annual growth rates in percent. The accounting data come from Compustat. Market returns data come from CRSP. The long-term forecasts come from IBES.

**Table 31**. Correlation of mean growth in operating income with long-term forecasts

Panel A: B/M quintiles

|  | whole sample | >1995 | <=1995 |
|---|---|---|---|
| 1 | 0.37*** | 0.38*** | 0.71*** |
| 2 | 0.28*** | 0.38*** | 0.30** |
| 3 | 0.15 | 0.16 | 0.54*** |
| 4 | 0.01 | 0.01 | 0.18 |
| 5 | -0.34*** | -0.17 | -0.50*** |

Panel B: Size quintiles

|   | whole sample | >1995 | <=1995 |
|---|---|---|---|
| 1 | 0.13 | 0.28** | -0.04 |
| 2 | 0.02 | 0.03 | 0.35** |
| 3 | 0.05 | 0.09 | 0.16 |
| 4 | 0.26*** | 0.26** | 0.46*** |
| 5 | 0.24*** | 0.24* | 0.24* |

Table 31: This table reports the correlations of the long-term growth forecasts of each book-to-market (Panel A) and size (Panel B) quintiles with the mean growth in operating income. Mean operating income growth is calculated over the previous four quarters. In all Panels, the data come with quarterly frequency. The forecasts were issued by analysts for the period 1981-2011; they are long-term forecasts, estimating an average annual growth of operating income for the following three to five years. The long-term growth forecasts are the forecasted annual growth rates in percent. Market returns data come from CRSP. The accounting data come from Compustat. The long-term forecasts come from IBES. One star indicates significance at the 10%, two stars at the 5% and three at the 1% level.

**Table 32**. Influence of forecasts of mean growth in operating income on the long-term forecasts of the B/M quintiles

Panel A: Whole Sample

|  | 1st B/M | 2nd B/M | 3rd B/M | 4th B/M | 5th B/M |
|---|---|---|---|---|---|
|  | Change in quintile aggregate estimates | | | | |
| Previous quarter Change of estimates | 0.07 | -0.17 | -0.01 | -0.08 | -0.05 |
| D.4-year Operating Income growth | 14.58*** | 11.26*** | 4.96* | -2.33 | -7.41** |
| LD.4-year Operating Income growth | -0.24 | -3.31 | 1.88 | 4.49 | 4.71 |
| Constant | -0.03 | -0.07 | 0.00 | 0.03 | 0.05 |
| N | 112 | 112 | 112 | 112 | 112 |
| R-sq | 0.155 | 0.112 | 0.048 | 0.022 | 0.109 |

\* $p<0.10$   \*\* $p<0.05$   \*\*\* $p<0.01$

Panel B: 1996-2011

|  | 1st B/M | 2nd B/M | 3rd B/M | 4th B/M | 5th B/M |
|---|---|---|---|---|---|
|  | Change in quintile aggregate estimates | | | | |
| Previous quarter Change of estimates | 0.10 | -0.10 | 0.02 | -0.06 | -0.04 |
| D.4-year Operating Income growth | 14.77** | 13.81*** | 2.80 | -3.56 | -9.41 |
| LD.4-year Operating Income growth | -1.72 | 2.41 | 4.26 | 4.93 | 8.20* |
| Constant | -0.06 | -0.05 | -0.04 | 0.00 | 0.04 |
| N | 62 | 62 | 62 | 62 | 62 |
| R-sq | 0.124 | 0.143 | 0.031 | 0.022 | 0.133 |

\* $p<0.10$   \*\* $p<0.05$   \*\*\* $p<0.01$

Panel C: 1981-1995

|  | 1st B/M | 2nd B/M | 3rd B/M | 4th B/M | 5th B/M |
|---|---|---|---|---|---|
|  | Change in quintile aggregate estimates | | | | |
| Previous quarter Change of estimates | -0.20 | -0.23 | -0.10 | -0.26** | -0.05 |
| D.4-year Operating Income growth | 13.73*** | 4.29 | 6.35* | 2.41 | -5.57** |
| LD.4-year Operating Income growth | 5.22** | -12.31** | 0.84 | 3.76 | -0.53 |
| Constant | 0.02 | -0.07 | 0.07 | 0.09 | 0.06 |
| N | 50 | 50 | 50 | 50 | 50 |
| R-sq | 0.396 | 0.279 | 0.118 | 0.097 | 0.192 |

\* p<0.10          \*\* p<0.05          \*\*\* p<0.01

Table 32: This table reports the regression results using the difference in the mean growth in operating incomes as independent variable and the difference in mean forecasted long-term growth rates of each B/M quintile as the dependent variable. Mean operating income growth is calculated over the previous four quarters. Panel A reports the results for the whole 1981-2011 period, Panel B for the 1996-2011 period, and Panel C for the 1981-1995 period. In all Panels, the data come with quarterly frequency. The forecasts were issued by analysts for the period 1981-2011; they are long-term forecasts, estimating an average annual growth of operating income for the next three to five years. The long-term growth forecasts are the forecasted annual growth rates in percent. The accounting data come from Compustat. Market returns data come from CRSP. The long-term forecasts come from IBES.

**Table 33**. Influence of forecasts of mean growth in operating income on the long-term forecasts of the size quintiles

Panel A: Whole Sample

|  | 1st Size | 2nd Size | 3rd Size | 4th Size | 5th Size |
|---|---|---|---|---|---|
|  | Change in quintile aggregate estimates | | | | |
| Previous quarter Change of estimates | -0.13* | -0.02 | 0.11 | -0.04 | -0.00 |
| D.4-year Operating Income growth | -4.79 | 1.67 | 13.62*** | 11.38* | 15.41** |
| LD.4-year Operating Income growth | 1.66 | -0.60 | -4.74 | 2.44 | -2.09 |
| Constant | -0.05 | 0.01 | -0.00 | -0.02 | -0.00 |
| N | 112 | 112 | 112 | 112 | 112 |
| R-sq | 0.066 | 0.003 | 0.153 | 0.071 | 0.114 |

\* p<0.10          \*\* p<0.05          \*\*\* p<0.01

Panel B: 1996-2011

|  | 1st Size | 2nd Size | 3rd Size | 4th Size | 5th Size |
|---|---|---|---|---|---|
|  | Change in quintile aggregate estimates | | | | |
| Previous quarter Change of estimates | 0.04 | 0.06 | 0.17** | 0.06 | -0.00 |
| D.4-year Operating Income growth | -11.23 | 5.85 | 17.17*** | 17.67 | 22.45* |
| LD.4-year Operating Income growth | 10.46** | -4.94 | -6.83 | 7.27 | -4.67 |
| Constant | -0.03 | -0.03 | -0.04 | -0.01 | -0.00 |
| N | 62 | 62 | 62 | 62 | 62 |
| R-sq | 0.104 | 0.033 | 0.204 | 0.150 | 0.134 |

* $p<0.10$  ** $p<0.05$  *** $p<0.01$

Panel C: 1981-1995

|  | 1st Size | 2nd Size | 3rd Size | 4th Size | 5th Size |
|---|---|---|---|---|---|
|  | Change in quintile aggregate estimates | | | | |
| Previous quarter Change of estimates | -0.27*** | -0.28** | -0.24* | -0.52*** | -0.07 |
| D.4-year Operating Income growth | -2.68 | -0.36 | 4.56 | 3.06 | 7.63* |
| LD.4-year Operating Income growth | -1.87 | 2.04 | -2.88 | -3.83 | -1.58 |
| Constant | -0.05 | 0.08 | 0.07 | -0.02 | 0.00 |
| N | 50 | 50 | 50 | 50 | 50 |
| R-sq | 0.205 | 0.098 | 0.119 | 0.477 | 0.226 |

* $p<0.10$  ** $p<0.05$  *** $p<0.01$

Table 33: This table reports the regression results using the difference in the mean growth in operating incomes as independent variable and the difference in mean long-term growth rates of each size quintile as the dependent variable. Mean operating income growth is calculated over the previous four quarters. Panel A reports the results for the whole 1981-2011 period, Panel B for the 1996-2011 period, and Panel C for the 1981-1995 period. In all Panels, the data come with quarterly frequency. The forecasts were issued by analysts for the period 1981-2011; they are long-term forecasts, estimating an average annual growth of operating income for the following three to five years. The long-term growth forecasts are the forecasted annual growth rates in percent. The accounting data come from Compustat. Market returns data come from CRSP. The long-term forecasts come from IBES.

**Table 34.** Interaction between quarterly forecasts and long-term growth forecasts

Panel A: Summary statistics of the quarterly forecast errors

|  | 10 | 25 | 50 | 75 | 90 | Mean | St. Dev. |
|---|---|---|---|---|---|---|---|
| Percentile | -0.066 | -0.01 | 0.01 | 0.036 | 0.09 | 0.001 | 0.4012 |

Panel B: Summary statistics of the long-term growth forecast changes between two subsequent quarters

|  | 10 | 25 | 50 | 75 | 90 | Mean | St. Dev. |
|---|---|---|---|---|---|---|---|
| Percentile | -2.5 | -0.5 | 0 | 0 | 1.5 | -0.26 | 3.98 |

Panel C: Summary statistics of long-term growth forecast changes between the current and the following quarters in the case of quarterly forecast errors of less than -0.066 dollars

|  | 10 | 25 | 50 | 75 | 90 | Mean | St. Dev. |
|---|---|---|---|---|---|---|---|
| Percentile | -2.5 | -0.5 | 0 | 0 | 1.5 | -0.2497 | 5.31 |

Panel D: Summary statistics of long-term growth forecast changes between the current and the following quarters in the case of quarterly forecast errors of more than 0.09 dollars

|  | 10 | 25 | 50 | 75 | 90 | Mean | St. Dev. |
|---|---|---|---|---|---|---|---|
| Percentile | -2 | -0.25 | 0 | 0 | 2 | -0.1158 | 4.38 |

Panel E: Summary statistics of long-term growth forecast changes between the current and the same quarters of the following year in the case of quarterly forecast errors of less than -0.066 dollars

|  | 10 | 25 | 50 | 75 | 90 | Mean | St. Dev. |
|---|---|---|---|---|---|---|---|
| Percentile | -5 | -2 | 0 | 0.5 | 2.5 | -0.864 | 5.66 |

Panel F: Summary statistics of long-term growth forecast changes between the current and the same quarters of the following year in the case of quarterly forecast errors of more than 0.09 cents

|  | 10 | 25 | 50 | 75 | 90 | Mean | St. Dev. |
|---|---|---|---|---|---|---|---|
| Percentile | -5 | -1.5 | 0 | 1 | 3 | -0.6367 | 6.03 |

Panel G: Summary statistics of quarterly forecast errors in the case of a change in long-term growth forecast between subsequent quarters that is smaller than -2.5%

|  | 10 | 25 | 50 | 75 | 90 | Mean | St. Dev. |
|---|---|---|---|---|---|---|---|
| Percentile | -0.07 | -0.012 | 0.01 | 0.035 | 0.09 | 0.00177 | 0.2234 |

Panel H: Summary statistics of quarterly forecast errors in the case of a change in long-term growth forecast between subsequent quarters that is greater than 1.5%

|  | 10 | 25 | 50 | 75 | 90 | Mean | St. Dev. |
|---|---|---|---|---|---|---|---|
| Percentile | -0.07 | -0.01 | 0.01 | 0.0333 | 0.0862 | -0.00086 | 0.4987 |

Table 34: This table reports summary statistics of the quarterly forecast errors, defined as the actual ESP minus the forecast (Panel A); summary statistics of long-term growth forecast changes between two subsequent quarters (Panel B); summary statistics of the long-term growth forecast changes between the current and the following quarters in the case that the quarterly forecast error belongs to the lowest decile of quarterly forecast errors (Panel C); summary statistics of the long-term growth forecast changes between the current and the next quarters in the case that the quarterly forecast error belongs to the top decile of quarterly forecast errors (Panel D); summary statistics of long-term growth forecast changes between the current and the same quarters of the following year in the case that the quarterly forecast error belongs to the lowest decile of quarterly forecast errors (Panel E); summary statistics of long-term growth forecast changes between the current and the same quarters of the following year in the case that the quarterly forecast error belongs to the top decile of quarterly forecast errors (Panel F); summary statistics of quarterly forecast errors in the case that the change in long-term growth forecast between subsequent quarters belongs to the lowest decile of changes in long-term growth forecasts (Panel G); and summary statistics of quarterly forecast errors in the case that the change in long-term growth forecast between subsequent quarters belongs to the top decile of changes in long-term growth forecasts (Panel H). The forecasts were issued by analysts for the period 1981-2011; they are long-term forecasts, estimating an average annual growth of operating income for the following three to five years. The long-term growth forecasts are the forecasted annual growth rates in percent. Market returns data come from CRSP. The long-term and quarterly forecasts come from IBES.

**Table 35**. Influence of mean quintile age and operating income growth on the long-term forecasts of the B/M quintiles

Panel A: Whole Sample

|  | 1st B/M | 2nd B/M | 3rd B/M | 4th B/M | 5th B/M |
|---|---|---|---|---|---|
|  | Change in quintile aggregate estimates | | | | |
| Previous quarter Change of estimates | -0.02 | -0.00 | 0.00 | 0.01 | 0.04 |
| D. quintile mean firm age | -0.23*** | -0.11*** | -0.05** | -0.12*** | -0.09*** |
| D.Four-year Operating Income growth | 6.75** | 7.05** | 5.05* | 1.03 | -3.23 |
| LD. quintile mean firm age | -0.00 | -0.00 | 0.01 | 0.01 | 0.00 |
| LD.Four-year Operating Income growth | -0.51 | 0.72 | -0.51 | 0.27 | 1.40 |
| Constant | 0.02 | 0.02 | 0.02 | 0.05 | 0.01 |
| N | 111 | 111 | 111 | 111 | 111 |
| R-sq | 0.495 | 0.363 | 0.166 | 0.383 | 0.230 |

\* p<0.10  \*\* p<0.05  \*\*\* p<0.01

Panel B: 1996-2011

|  | 1st B/M | 2nd B/M | 3rd B/M | 4th B/M | 5th B/M |
|---|---|---|---|---|---|
|  | Change in quintile aggregate estimates | | | | |
| Previous quarter Change of estimates | -0.03 | -0.01 | -0.01 | 0.02 | 0.07 |
| D. quintile mean firm age | -0.25*** | -0.09 | -0.04 | -0.14*** | -0.13*** |
| D.Four-year Operating Income growth | 7.25* | 9.05** | 2.43 | 0.55 | -1.92 |
| LD. quintile mean firm age | -0.01 | -0.01 | 0.00 | 0.02 | 0.01 |
| LD.Four-year Operating Income growth | -1.15 | 1.51 | -0.46 | -0.20 | 2.60 |
| Constant | 0.08 | 0.01 | -0.01 | 0.09 | 0.02 |
| N | 62 | 62 | 62 | 62 | 62 |
| R-sq | 0.467 | 0.278 | 0.134 | 0.402 | 0.257 |

\* $p<0.10$  \*\* $p<0.05$  \*\*\* $p<0.01$

Panel C: 1981-1995

|  | 1st B/M | 2nd B/M | 3rd B/M | 4th B/M | 5th B/M |
|---|---|---|---|---|---|
|  | Change in quintile aggregate estimates | | | | |
| Previous quarter Change of estimates | -0.03 | -0.01 | 0.11 | -0.05 | -0.01 |
| D. quintile mean firm age | -0.19*** | -0.16*** | -0.07 | -0.09*** | -0.05* |
| D.Four-year Operating Income growth | 6.75** | 1.99 | 7.77* | 4.57 | -3.63* |
| LD. quintile mean firm age | -0.01 | 0.01 | 0.01 | -0.00 | -0.00 |
| LD.Four-year Operating Income growth | 0.04 | -0.10 | -0.14 | 0.40 | 0.22 |
| Constant | -0.03 | 0.00 | 0.05 | 0.03 | 0.01 |
| N | 49 | 49 | 49 | 49 | 49 |
| R-sq | 0.671 | 0.710 | 0.254 | 0.466 | 0.281 |

\* $p<0.10$  \*\* $p<0.05$  \*\*\* $p<0.01$

Table 35: This table reports the regression results using the difference in mean quintile firm age and difference in mean growth in operating income as the independent variables and the difference in the mean long-term forecast of each B/M quintile as the dependent variable. Firm age is defined as the number of quarters the firm is publicly traded. Mean operating income growth is calculated over the previous four quarters. Panel A reports the results for the whole 1981-2011 period, Panel B for the 1996-2011 period, and Panel C for the 1981-1995 period. In all Panels, the data come with quarterly frequency. The long-term forecasts were issued by analysts for the period 1981-2011; they estimate an average annual growth of operating income for the following three to five years. The long-term growth forecasts are the forecasted annual growth rates in percent. The accounting data come from Compustat. Market returns data come from CRSP. The long-term forecasts come from IBES.

**Table 36**. Influence of mean quintile age and operating income growth on the long-term forecasts of the size quintiles

Panel A: Whole Sample

|  | 1st Size | 2nd Size | 3rd Size | 4th Size | 5th Size |
|---|---|---|---|---|---|
|  | Change in quintile aggregate estimates | | | | |
| Previous quarter Change of estimates | -0.03 | -0.13* | 0.01 | 0.08 | -0.05 |
| D. quintile mean firm age | -0.13*** | -0.07** | -0.09** | -0.10** | -0.12*** |
| D.Four-year Operating Income growth | -1.60 | 1.82 | 9.23*** | 11.73** | 8.26*** |
| LD. quintile mean firm age | -0.00 | -0.02* | 0.00 | 0.01 | 0.00 |
| LD.Four-year Operating Income growth | 4.50 | 1.55 | 0.58 | -6.65 | 2.94 |
| Constant | -0.02 | 0.01 | 0.01 | 0.03 | 0.02 |
| N | 111 | 111 | 111 | 111 | 111 |
| R-sq | 0.164 | 0.099 | 0.254 | 0.398 | 0.585 |

\* $p<0.10$    \*\* $p<0.05$    \*\*\* $p<0.01$

Panel B: 1996-2011

|  | 1st Size | 2nd Size | 3rd Size | 4th Size | 5th Size |
|---|---|---|---|---|---|
|  | Change in quintile aggregate estimates | | | | |
| Previous quarter Change of estimates | -0.07 | -0.15* | -0.03 | 0.07 | -0.06 |
| D. quintile mean firm age | -0.16 | -0.11 | -0.17** | -0.17*** | -0.14*** |
| D.Four-year Operating Income growth | -2.74 | 2.50 | 11.17*** | 20.93** | 11.34*** |
| LD. quintile mean firm age | -0.01 | -0.02 | -0.01 | -0.00 | 0.00 |
| LD.Four-year Operating Income growth | 1.25 | 2.72 | -2.82 | -17.40* | 4.65 |
| Constant | 0.05 | 0.03 | 0.07 | 0.06 | -0.01 |
| N | 62 | 62 | 62 | 62 | 62 |
| R-sq | 0.143 | 0.095 | 0.367 | 0.643 | 0.639 |

\* $p<0.10$    \*\* $p<0.05$    \*\*\* $p<0.01$

Panel C: 1981-1995

|  | 1st Size | 2nd Size | 3rd Size | 4th Size | 5th Size |
|---|---|---|---|---|---|
|  | Change in quintile aggregate estimates | | | | |
| Previous quarter Change of estimates | -0.05 | 0.05 | -0.16 | -0.05 | -0.03 |
| D. quintile mean firm age | -0.12*** | -0.06** | -0.04* | -0.01 | -0.05*** |
| D.Four-year Operating Income growth | -0.20 | 0.17 | 0.90 | 3.48 | 4.86 |
| LD. quintile mean firm age | 0.02 | -0.02 | 0.00 | -0.00 | -0.00 |
| LD.Four-year Operating Income growth | 8.04 | -1.61 | 3.69 | 1.76 | 1.28 |
| Constant | -0.09 | 0.01 | 0.03 | 0.03 | 0.03 |
| N | 49 | 49 | 49 | 49 | 49 |
| R-sq | 0.310 | 0.172 | 0.209 | 0.138 | 0.458 |

\* p<0.10      \*\* p<0.05      \*\*\* p<0.01

Table 36: This table reports the regression results using the difference in mean quintile firm age and mean growth in operating income as the independent variables and the difference in the mean long-term forecast of each size quintile as the dependent variable. Firm age is defined as the number of quarters the firm is publicly traded. Mean operating income growth is calculated over the previous four quarters. Panel A reports the results for the whole 1981-2011 period, Panel B for the 1996-2011 period, and Panel C for the 1981-1995 period. In all Panels, the data come with quarterly frequency. The long-term forecasts were issued by analysts for the period 1981-2011; they estimate an average annual growth of operating income for the following three to five years. The long-term growth forecasts are the forecasted annual growth rates in percent. The accounting data come from Compustat. Market returns data come from CRSP. The long-term forecasts come from IBES.

**Table 37**. Influence of IPOs and TFP growth on long-term estimates, mean quintile age, and mean quintile operating income of B/M quintiles

Panel A: IPOs

|  | 1st B/M | 2nd B/M | 3rd B/M | 4th B/M | 5th B/M |
|---|---|---|---|---|---|
| Difference in mean long-term estimates growth | 0.52*** | 0.11 | 0.24** | 0.14 | 0.05 |
| Difference in mean age growth | -0.74* | -0.31 | -1.02 | -0.64 | 0.59 |
| Difference in mean operating income growth | 0.00 | 0.00 | 0.01 | -0.00 | 0.01 |

Panel B: TFP growth

|  | 1st B/M | 2nd B/M | 3rd B/M | 4th B/M | 5th B/M |
|---|---|---|---|---|---|
| Difference in mean long-term estimates growth | 0.34** | 0.09 | 0.05 | -0.15 | -0.16 |
| Difference in mean age growth | -0.34 | 0.76* | 1.32 | 0.66 | 1.13 |
| Difference in mean operating income growth | 0.01* | 0.02*** | 0.02*** | 0.01*** | 0.03*** |

Table 37: This table reports the difference in means of periods of low and high IPO activity (Panel A) and low and high TFP (Panel B). The whole period 1981-2011 is divided into two categories for each B/M quintile. The first category includes quarters with more than the median number of IPOs. For each category, I calculate the mean of change in long-term forecasts, change in quintile age, and change in quintile operating income; then I calculate the difference in the means between the two categories for each quintile. I repeat the same procedure for the TFP growth. The long-term forecasts were issued by analysts for the period 1981-2011; they estimate an average annual growth of operating income for the following three to five years. The long-term growth forecasts are the forecasted annual growth rates in percent. The accounting data come from Compustat. Market returns data come from CRSP. The long-term forecasts come from IBES. One star indicates significance at the 10%, two stars at the 5% and three at the 1% level.

**Table 38**. Influence of IPOs and TFP growth on long-term estimates, mean quintile age, and mean quintile operating income of size quintiles

Panel A: IPOs

|  | 1st Size | 2nd Size | 3rd Size | 4th Size | 5th Size |
|---|---|---|---|---|---|
| Difference in mean long-term estimates growth | 0.09 | 0.23 | 0.40*** | 0.37* | 0.37** |
| Difference in mean age growth | 0.03 | -0.78 | -0.47 | -1.90* | -1.56* |
| Difference in mean operating income growth | 0.00 | -0.00 | 0.00 | 0.00 | 0.00 |

Panel B: TFP growth

|  | 1st Size | 2nd Size | 3rd Size | 4th Size | 5th Size |
|---|---|---|---|---|---|
| Difference in mean long-term estimates growth | 0.05 | 0.10 | 0.23 | 0.30 | 0.38** |
| Difference in mean age growth | 0.74 | 0.67 | 0.84 | 0.91 | -1.13 |
| Difference in mean operating income growth | 0.02*** | 0.01** | 0.01** | 0.01** | 0.01*** |

Table 38: This table reports the difference in means of periods of low and high IPO activity (Panel A) and low and high TFP (Panel B). The whole period 1981-2011 is divided into two categories for each size quintile. The first category includes quarters with more the median number of IPOs. For each category, I calculate the mean of change in long-term forecasts, change in quintile age, and change in quintile operating income; then I calculate the difference in the means between the two categories for each quintile. I repeat the same procedure for the TFP growth. The long-term forecasts were issued by analysts for the period 1981-2011; they estimate an average annual growth of operating income for the following three to five years. The long-term growth forecasts are the forecasted annual growth rates in percent. The accounting data come from Compustat. Market returns data come from CRSP. The long-term forecasts come from IBES. One star indicates significance at the 10%, two stars at the 5% and three at the 1% level.

**Figure 1.** Number of firms which have the long-term growth forecasts in IBES

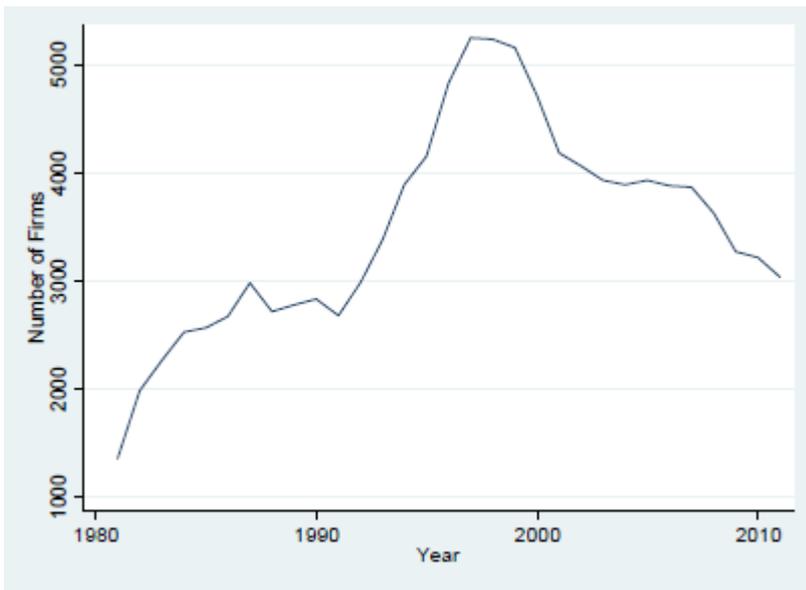

Figure 1: This figure presents the number of firms present at IBES every year.

**Figure 2.** Number of long-term growth forecasts in IBES

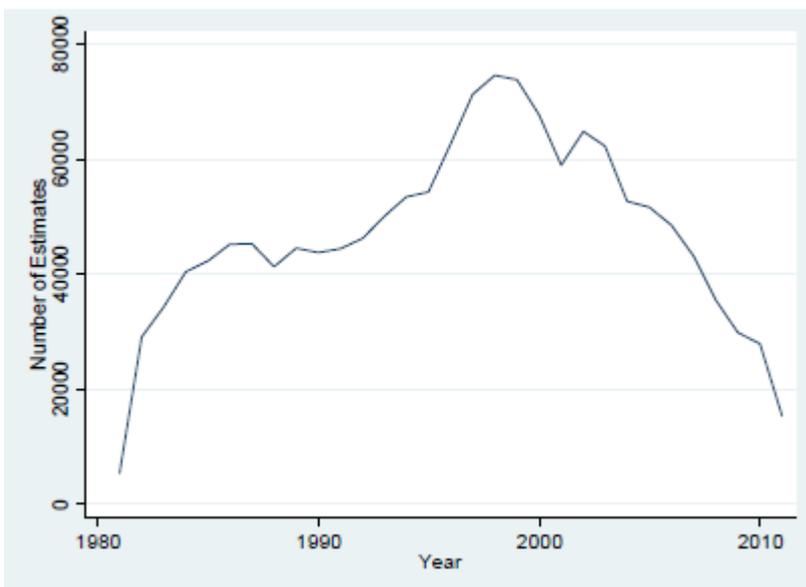

Figure 2: This figure presents the number of long-term forecasts per year in IBES.

**Figure 3.** Number of firms in each book-to-market quintile

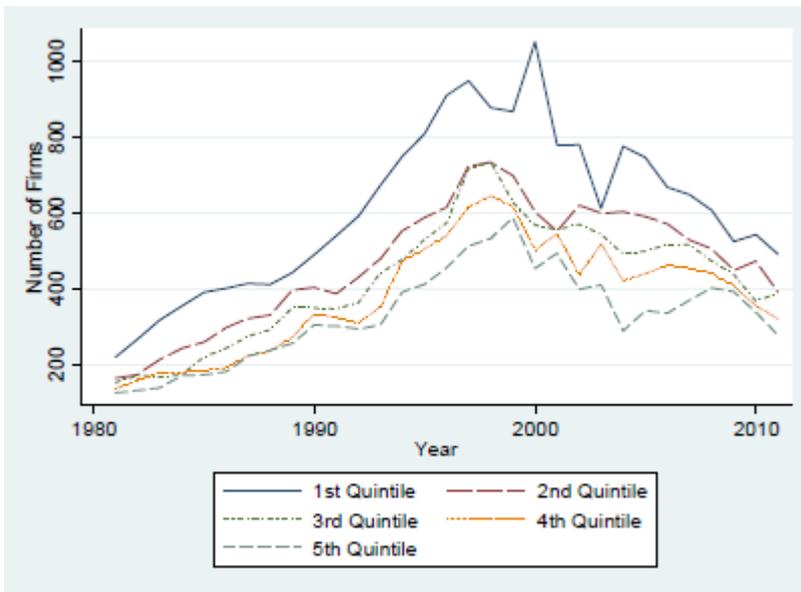

Figure 3: This figure presents the number of firms in each book-to-market quintile for which long-term forecasts are available in IBES. The accounting data come from Compustat, and the stock price data come from CRSP.

**Figure 4.** Number of firms in each size quintile

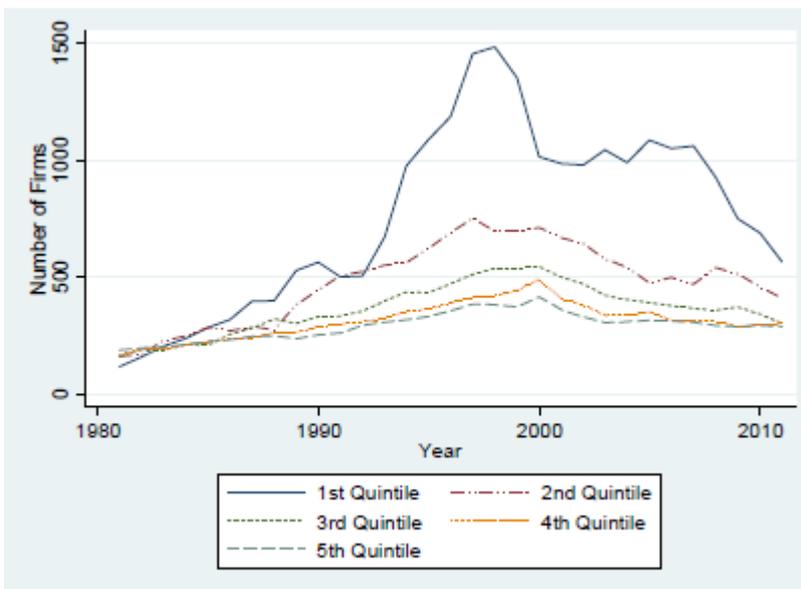

Figure 4: This figure presents the number of firms in each size quintile for which long-term forecasts are available in IBES. The stock price data come from CRSP.

**Figure 5**. Operating income for Microsoft

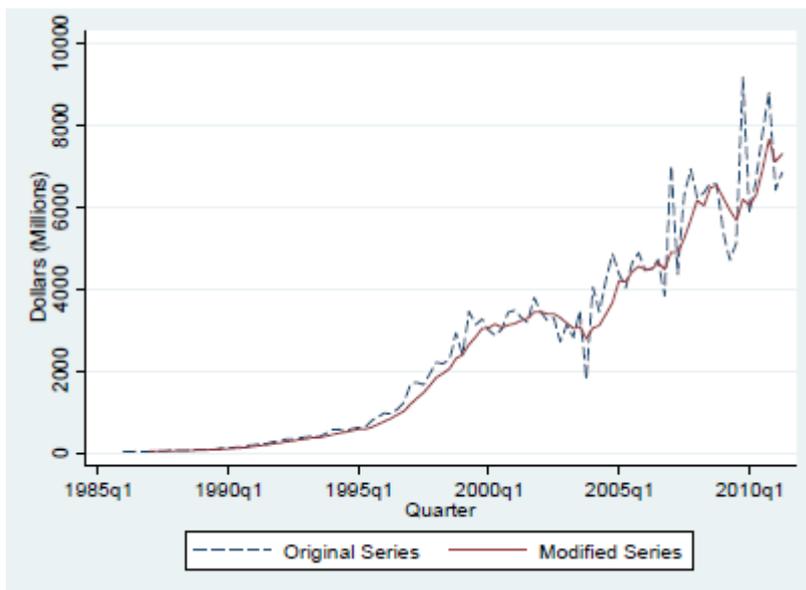

Figure 5: This figure presents the original and the smoothed series of operating income for Microsoft. The original data were smoothed by using a moving average of the current and the previous four quarters of the firm's operating income. The accounting data come from Compustat.

**Figure 6**. Long-term forecasts of B/M quintiles

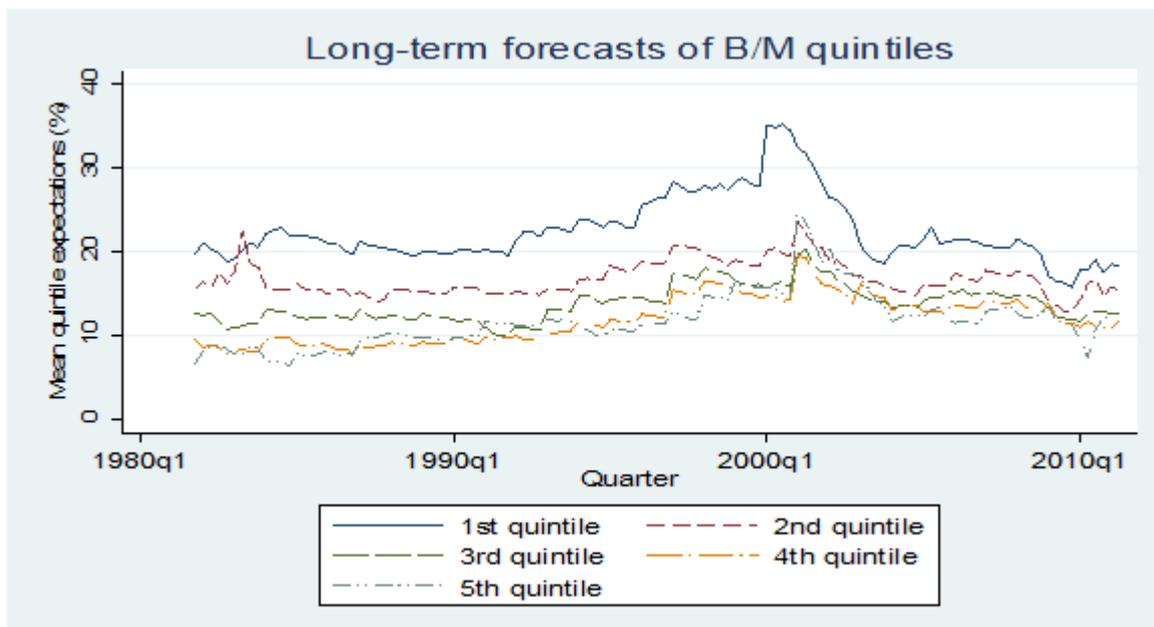

Figure 6: This figure presents the long-term forecasts for each book-to-market quintile. The long-term growth forecasts are the forecasted annual growth rates of operating income in percent. The accounting data come from Compustat. Stock data come from CRSP. The long-term forecasts come from IBES.

**Figure 7**. Relative errors for B/M quintiles

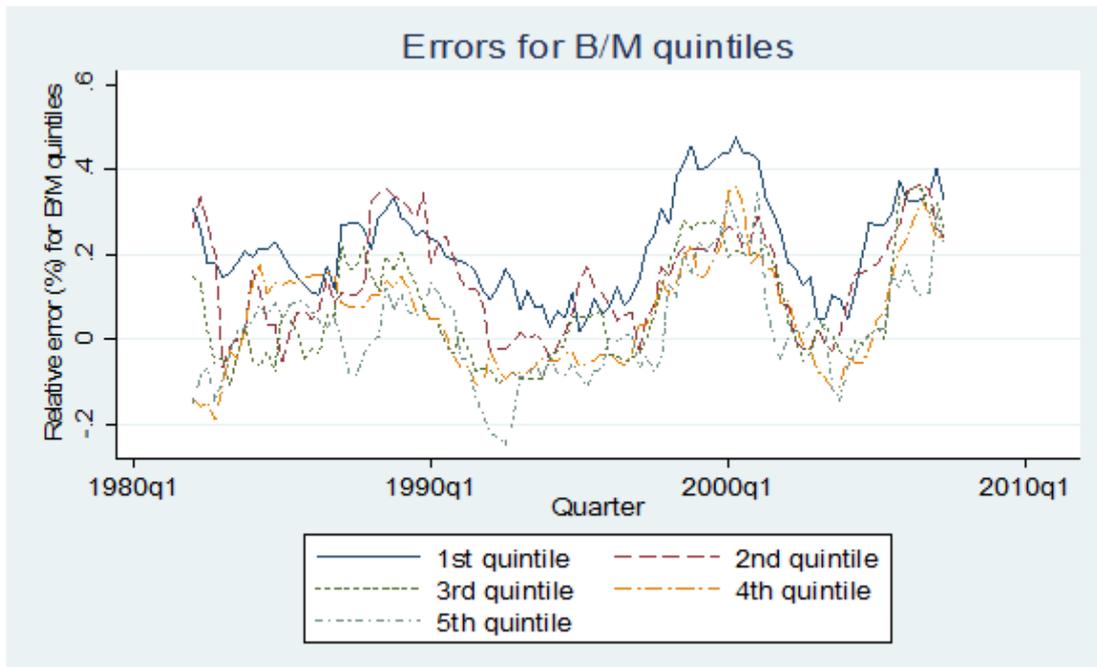

Figure 7: This figure presents the relative errors for each book-to-market quintile. The relative error is defined as the four-years-ahead forecasted operating income minus the actual operating income four years later, divided by the actual operating income four years later. The long-term growth forecasts are the forecasted annual growth rates of operating income in percent. The accounting data come from Compustat. Stock data come from CRSP. The long-term forecasts come from IBES.

**Figure 8**. Long-term forecasts of size quintiles

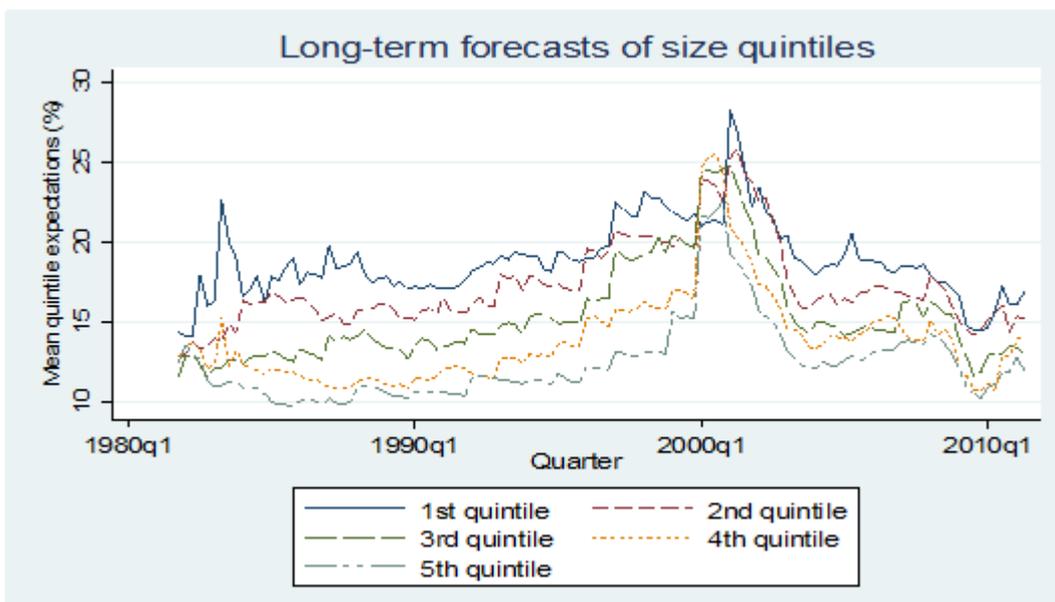

Figure 8: This figure presents the long-term forecasts for each size quintile. The long-term growth forecasts are the forecasted annual growth rates of operating income in percent. The accounting data come from Compustat. Stock data come from CRSP. The long-term forecasts come from IBES.

**Figure 9**. Relative errors for size quintiles

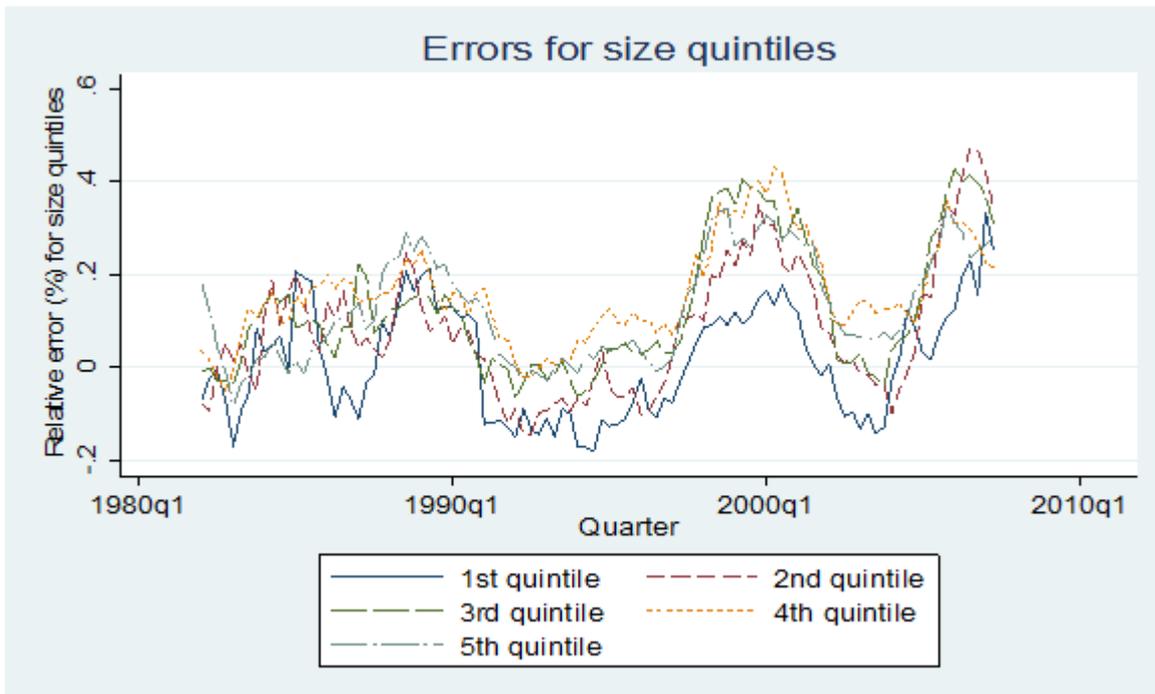

Figure 9: This figure presents the relative errors for each size quintile. The relative error is defined as the four-years-ahead forecasted operating income minus the actual operating income four years later, divided by the actual operating income four years later. The long-term growth forecasts are the forecasted annual growth rates of operating income in percent. The accounting data come from Compustat. Stock data come from CRSP. The long-term forecasts come from IBES.

**Figure 10**. GDP growth forecasts

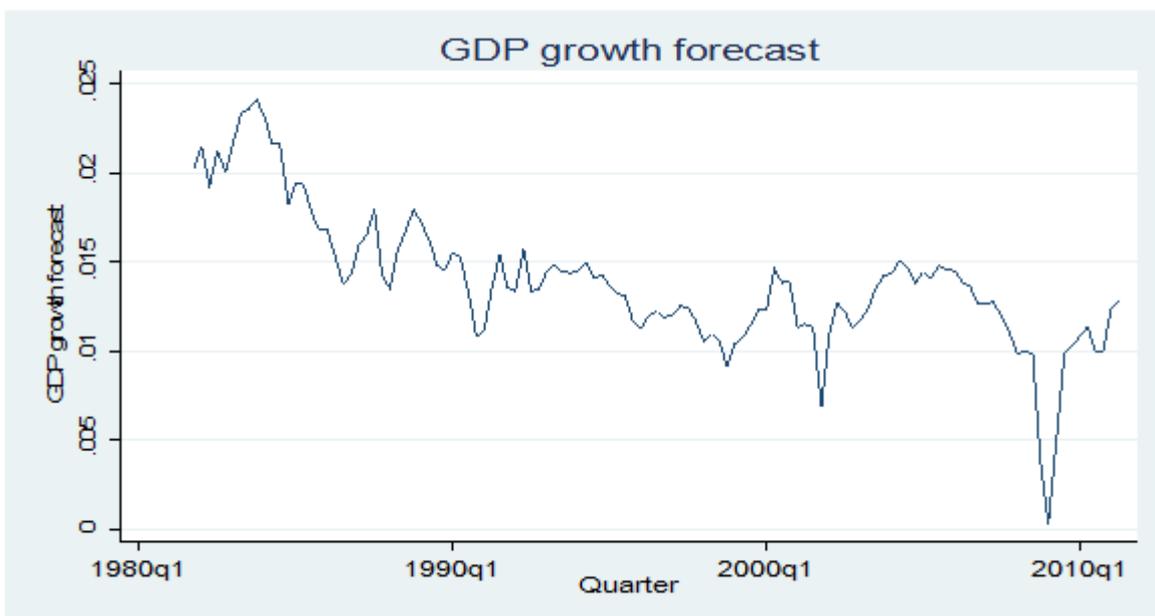

Figure 10: This figure presents the one-year-ahead forecasted GDP growth. The GDP forecasts come from the Survey of Professional Forecasters.

**Figure 11**. Corporate profits forecasts

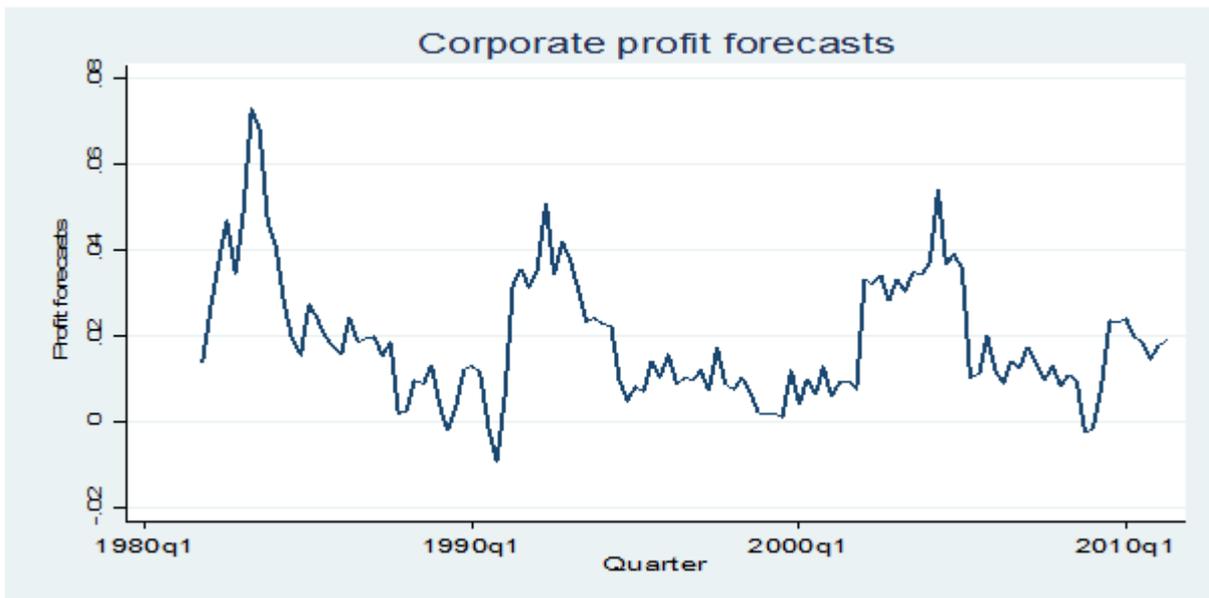

Figure 11: This figure presents the four-quarters-ahead corporate profits growth forecasts. The corporate profits forecasts come from the Survey of Professional Forecasters.

**Figure 12**. Comparison of long-term forecasts for the first B/M quintile, with and without the new firms

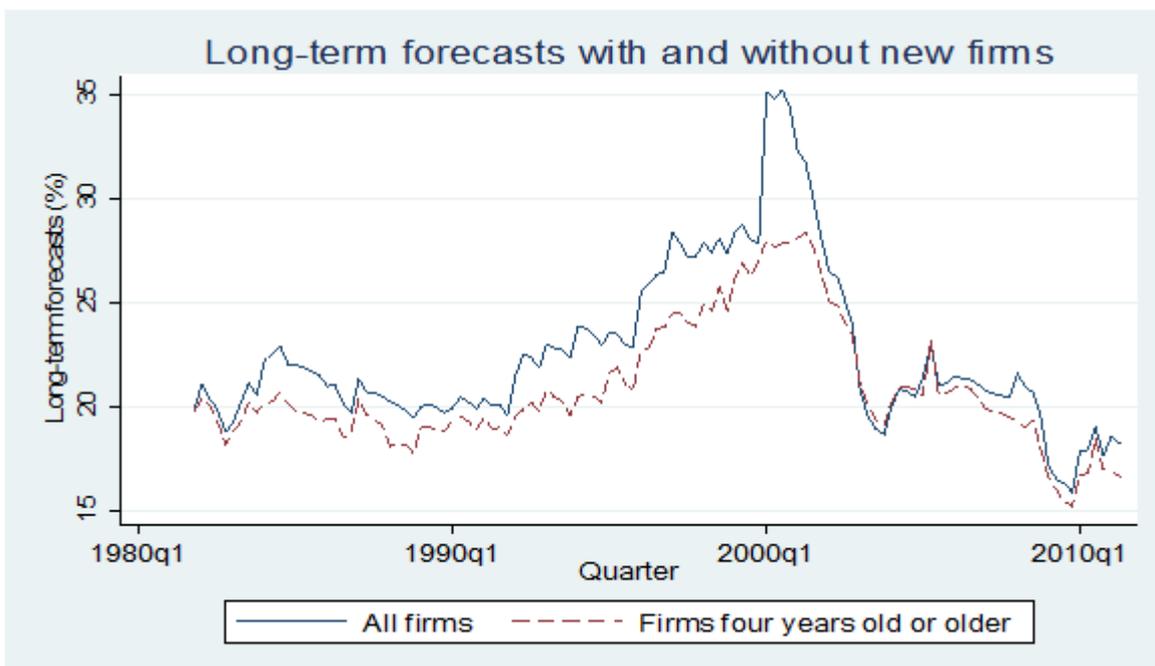

Figure 12: This figure presents the long-term forecasts for the first B/M quintile. The "All firms" line includes all firms that belong to the quintile. The "firms four years old or older" line includes only the firms that trade in a stock market for four years or more. The long-term growth forecasts are the forecasted annual growth rates of operating income in percent. The accounting data come from Compustat. Stock data come from CRSP. The long-term forecasts come from IBES.

**Figure 13**. Comparison of long-term forecasts for the third B/M quintile, with and without the new firms

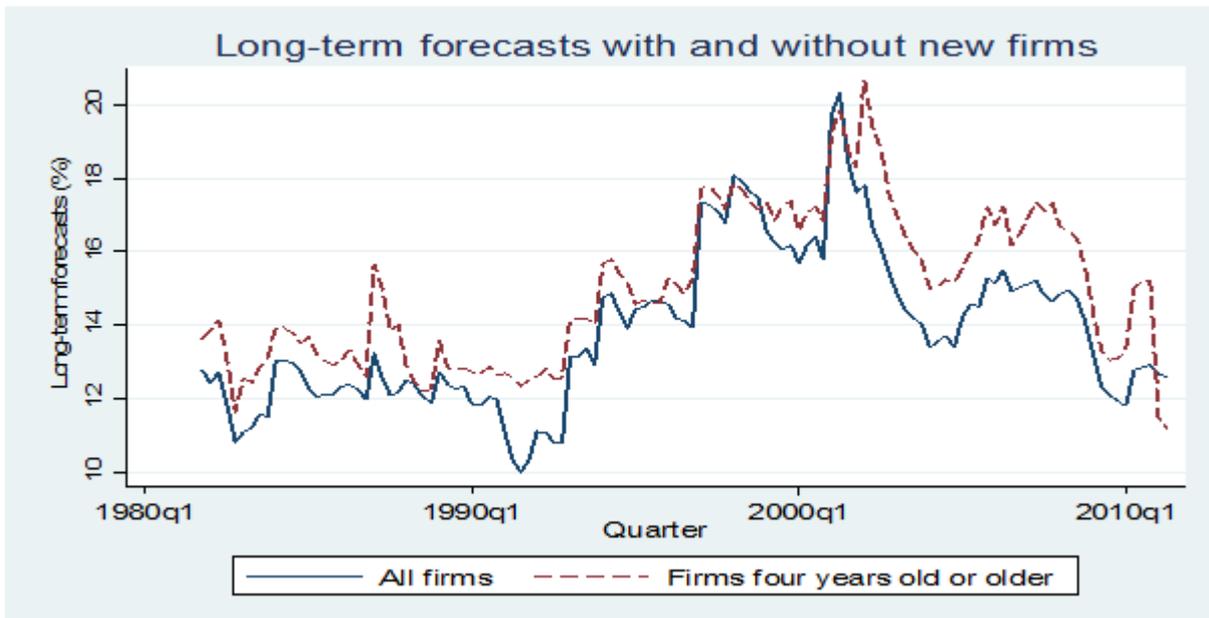

Figure 13: This figure presents the long-term forecasts for the third B/M quintile. The "All firms" line includes all firms that belong to the quintile. The "firms four years old or older" line includes only the firms that trade in a stock market for four years or more. The long-term growth forecasts are the forecasted annual growth rates of operating income in percent. The accounting data come from Compustat. Stock data come from CRSP. The long-term forecasts come from IBES.

**Figure 14**. Comparison of long-term forecasts for the fifth B/M quintile, with and without the new firms

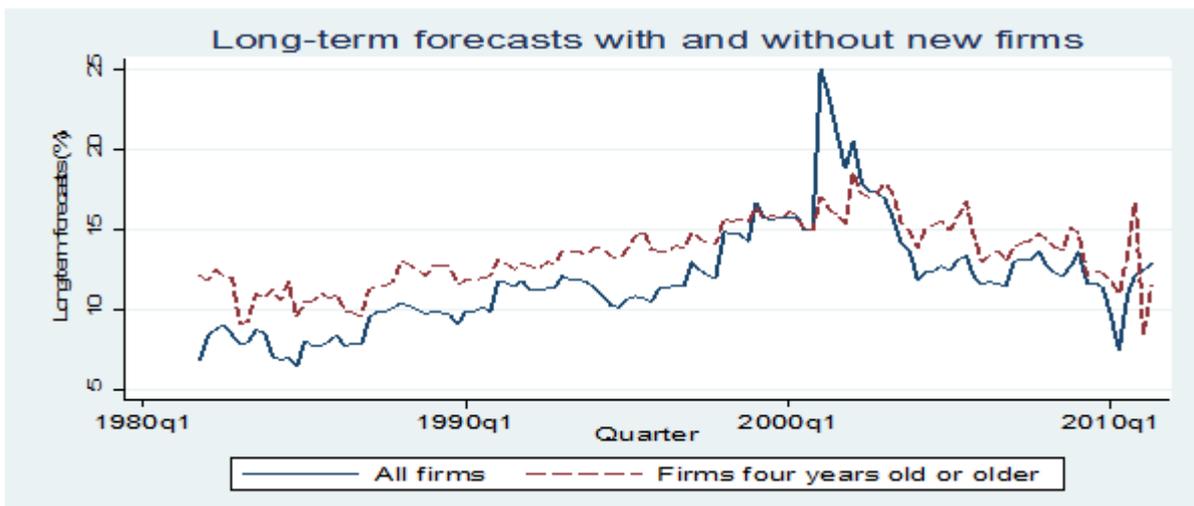

Figure 14: This figure presents the long-term forecasts for the 5$^{th}$ B/M quintile. The "All firms" line includes all firms that belong to the quintile. The "firms four years old or older" line includes only the firms that trade in a stock market for four years or more. The long-term growth forecasts are the forecasted annual growth rates of operating income in percent. The accounting data come from Compustat. Stock data come from CRSP. The long-term forecasts come from IBES.

**Figure 15**. Comparison of long-term forecasts for the 1st size quintile, with and without the new firms

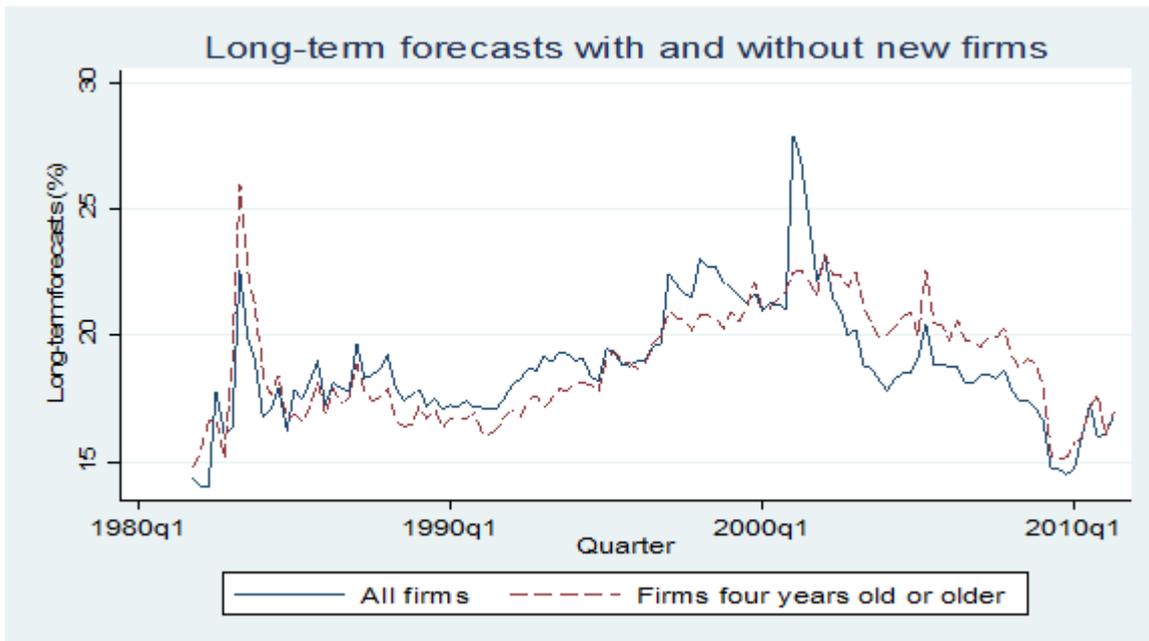

Figure 15: This figure presents the long-term forecasts for the 1st size quintile. The "All firms" line includes all firms that belong to the quintile. The "firms four years old or older" line includes only the firms that trade in a stock market for four years or more. The long-term growth forecasts are the forecasted annual growth rates of operating income in percent. The accounting data come from Compustat. Stock data come from CRSP. The long-term forecasts come from IBES.

**Figure 16**. Comparison of long-term forecasts for the 3rd size quintile, with and without the new firms

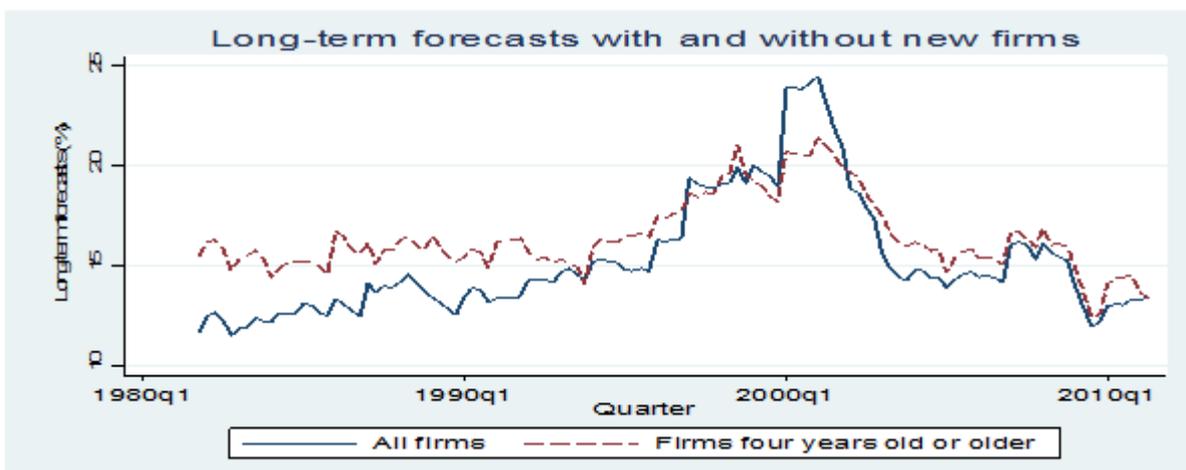

Figure 16: This figure presents the long-term forecasts for the 3rd size quintile. The "All firms" line includes all firms that belong to the quintile. The "firms four years old or older" line includes only the firms that trade in a stock market for four years or more. The long-term growth forecasts are the forecasted annual growth rates of operating income in percent. The accounting data come from Compustat. Stock data come from CRSP. The long-term forecasts come from IBES.

**Figure 17**. Comparison of long-term forecasts for the 5<sup>th</sup> size quintile, with and without the new firms

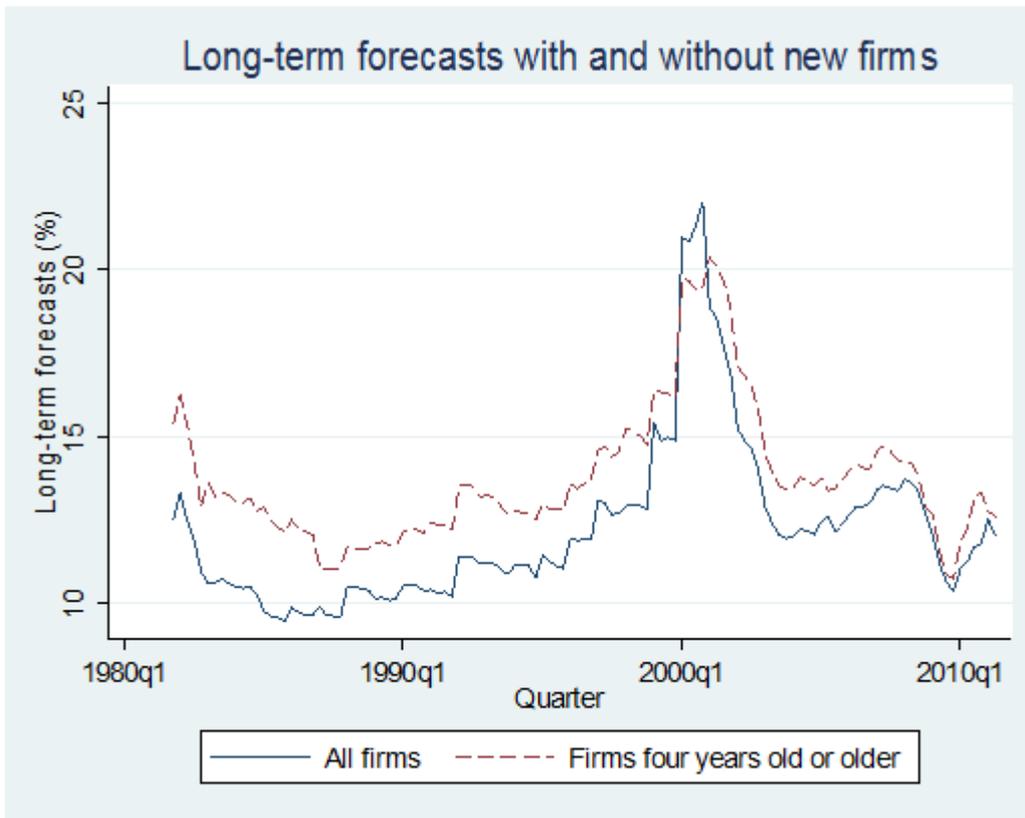

Figure 17: This figure presents the long-term forecasts for the 5<sup>th</sup> size quintile. The "All firms" line includes all firms that belong to the quintile. The "firms four years old or older" line includes only the firms that trade in a stock market for four years or more. The long-term growth forecasts are the forecasted annual growth rates of operating income in percent. The accounting data come from Compustat. Stock data come from CRSP. The long-term forecasts come from IBES.